\documentclass[number,preprint,3p]{elsarticle}

\usepackage{multirow}
\usepackage{color}
\usepackage{graphicx}
\usepackage{subcaption}
\usepackage{algorithm}
\usepackage{bm}
\usepackage[colorlinks]{hyperref}
\usepackage{amssymb}
\usepackage{amsthm}
\usepackage{amsmath}
\usepackage{booktabs}
\usepackage{url}
\usepackage{scrextend}
\usepackage{epstopdf}
\usepackage{float}
\usepackage{tablefootnote}
\usepackage{mathtools}
\usepackage{soul}
\usepackage[perpage]{footmisc}
\usepackage{lineno}
\makeatletter
\newenvironment{breakablealgorithm}
  {
   \begin{center}
     \refstepcounter{algorithm}
     \hrule height.8pt depth0pt \kern2pt
     \renewcommand{\caption}[2][\relax]{
       {\raggedright\textbf{\ALG@name~\thealgorithm} ##2\par}%
       \ifx\relax##1\relax 
         \addcontentsline{loa}{algorithm}{\protect\numberline{\thealgorithm}##2}%
       \else 
         \addcontentsline{loa}{algorithm}{\protect\numberline{\thealgorithm}##1}%
       \fi
       \kern2pt\hrule\kern2pt
     }
  }{
     \kern2pt\hrule\relax
   \end{center}
  }
\makeatother
\makeatletter
\gdef\urlauthor#1#2{\g@addto@macro\@elsuads{\let\corref\@gobble%
     \def\@@tmp{#1}\raggedright\eadsep
     {\ttfamily\url{\expandafter\strip@prefix\meaning\@@tmp}}\space(#2)%
     \def\eadsep{\unskip,\space}}%
}
\gdef\emailauthor#1#2{\stepcounter{ead}%
     \g@addto@macro\@elseads{\raggedright%
      \let\corref\@gobble\def\@@tmp{#1}%
      \eadsep{\ttfamily\href{mailto:\expandafter\strip@prefix\meaning\@@tmp}{\expandafter\strip@prefix\meaning\@@tmp}}
      (#2)\def\eadsep{\unskip,\space}}%
}
\makeatother

\def\r{\mathbb{R}}
\def\rn{\mathbb{R}^n}

\newcommand{\vect}[1]{\boldsymbol{#1}}

\makeatletter
\let\@afterindenttrue\@afterindentfalse
\makeatother

\biboptions{numbers,sort&compress,square}
\journal{Earthquake Engineering and Structural Dynamics}
\begin{document}
\begin{frontmatter}
\renewcommand{\thefootnote}{\fnsymbol{footnote}}



\title{Uncertainty quantification for seismic response using \\dimensionality reduction-based stochastic simulator}

 \author[1]{Jungho Kim}
 \author[1]{Ziqi Wang\corref{cor1}}
\ead{ziqiwang@berkeley.edu}  
         \cortext[cor1]{Corresponding author}
 \address[1]{Department of Civil and Environmental Engineering, University of California, Berkeley, United States}
\begin{abstract}
This paper introduces a stochastic simulator for seismic uncertainty quantification, which is crucial for performance-based earthquake engineering. The proposed simulator extends the recently developed dimensionality reduction-based surrogate modeling method (DR-SM) to address high-dimensional ground motion uncertainties and the high computational demands associated with nonlinear response history analyses. By integrating physics-based dimensionality reduction with multivariate conditional distribution models, the proposed simulator efficiently propagates seismic input into multivariate response quantities of interest. The simulator can incorporate both aleatory and epistemic uncertainties and does not assume distribution models for the seismic responses. The method is demonstrated through three finite element building models subjected to synthetic and recorded ground motions. The proposed method effectively predicts multivariate seismic responses and quantifies uncertainties, including correlations among responses.
\end{abstract}
\end{frontmatter}
\renewcommand{\thefootnote}{\fnsymbol{footnote}}

\section{Introduction}

\noindent Quantifying the variability in seismic responses, propagated from diverse sources of uncertainty that affect structural performance, is crucial for performance-based earthquake engineering (PBEE) \cite{porter2003overview,moehle2004framework,gunay2013peer} and seismic risk assessment \cite{araya1988seismic,ellingwood2009quantifying,baker2021seismic,kyprioti2022addressing}. Uncertainty quantification (UQ) for seismic response entails addressing high-dimensional uncertainties from seismic hazard models, structural systems, and the inherent randomness in ground motions. A primary computational challenge in seismic UQ is the intensive computational demand imposed by nonlinear response history analysis (NLRHA) of structural models. Despite computational advances, the cost of high-fidelity simulations under extensive ground motion datasets remains prohibitively high, underscoring the need for improved efficiency in uncertainty propagation for seismic responses.

Recent advancements in computational UQ encompass a spectrum of methods, including efficient time series analysis \cite{enescu2006wavelet}, advanced simulation techniques \cite{jayaram2010efficient,wang2019hamiltonian,patsialis2021multi,rudman2024assessment}, reduced-order modeling \cite{patsialis2020reduced,ruggieri2022reduced}, statistical linearization \cite{broccardo2016multicomponent,wang2017equivalent,wang2024optimized}, and surrogate modeling \cite{kyprioti2021kriging,kim2021quantile,zhong2023surrogate,kim2024active}. In particular, for analysis scenarios requiring repeated NLRHAs of complex structural models, surrogate modeling emerges as a key strategy. Popular surrogate models include Kriging/Gaussian process \cite{rasmussen2003gaussian,gidaris2015kriging,kim2021reliability}, polynomial chaos expansion \cite{sudret2008global,papaioannou2019pls,zhu2023seismic}, and neural networks \cite{gurney2018introduction,tripathy2018deep,kim2020probabilistic}. Despite their efficiency, these models face significant challenges due to high-dimensional uncertainties and the complex uncertainty propagation through NLRHA. To address these issues, the authors recently developed a method that extracts surrogate models from the outcomes of dimensionality reduction, termed the dimensionality reduction-based surrogate model method (DR-SM) \cite{kim2024dimensionality}. While DR-SM proves effective even with high-dimensional inputs, its applications to seismic UQ require further development because (i) traditional, non-intrusive dimensionality reduction algorithms are not optimized for this application, and (ii) DR-SM is primarily designed for single-output analyses; its capability to predict multiple response quantities and their correlations needs further investigation.

Building on DR-SM, this paper introduces a stochastic simulator tailored for seismic UQ analysis with multiple response quantities. This simulator integrates physics-based dimensionality reduction with a generic dimensionality reduction algorithm applied to an augmented input-output space. The method employs a mixture-based distribution model to fit multivariate conditional distributions within the reduced feature space, thereby accommodating multiple responses. The stochastic simulator is extracted from simulating a transition kernel that involves iterative dimensionality reduction and conditional distribution sampling. Due to the properties of this iterative process, potential Gaussian assumptions in feature space modeling do not necessarily lead to Gaussian responses. Consequently, the proposed approach can predict non-Gaussian seismic responses from high-dimensional inputs, efficiently propagating both epistemic and aleatory uncertainties.

Importantly, the proposed approach does not necessitate the probabilistic distribution of response quantities, which are typically assumed to be lognormal variables in the conventional methodologies \cite{kyprioti2021kriging,kim2023estimation,yi2024stochastic}. Such methods often fail to accurately reflect the actual response distribution. Moreover, in contrast to numerous existing studies, this work effectively captures the correlation structures among various response quantities, such as the interdependence between peak story drift ratios at different building heights, which is crucial for detailed damage and loss assessments. Additionally, this method is applicable to both synthetically generated stochastic ground motions and real ground motion records, making it useful under various seismic UQ practices.

This paper first states the problem and provides an overview of the DR-SM method \cite{kim2024dimensionality} in Section \ref{Overview}. Section \ref{Proposed} details the proposed approach, including (i) physics-based dimensionality reduction, (ii) multivariate distribution model, (iii) determination of the reduced dimensionality, and (iv) the algorithm of the proposed simulator. Section \ref{Examples} demonstrates the performance of the proposed method with numerical examples. The paper concludes with a summary in Section \ref{Conclusion}.

\section{Overview of seismic response UQ using surrogate models} \label{Overview}

\subsection{Problem formulation and sources of uncertainty} \label{Problem}

\noindent The quantification of seismic response uncertainties is facilitated using NLRHA. The input random variables for NLRHA, $\vect{X}\in\r^{n}$, consist of three  categories: (i) seismic hazard characteristics $\vect{X}_{h}\in\r^{n_{h}}$, including the moment magnitude of the seismic event and the rupture distance; (ii) excitation sequences $\vect{X}_{w}\in\r^{n_{w}}$, representing aleatory uncertainties within the ground motions, which encompass white noise sequences utilized in the stochastic ground motions model (SGMM); and (iii) structural parameters $\vect{X}_{s}\in\r^{n_{s}}$, such as material properties and damping ratios. Collectively, these inputs form a parameter set $\vect{X}=\left[\vect{X}_{h},\vect{X}_{w},\vect{X}_{s}\right]$ with a total dimension of $n=n_{h}+n_{w}+n_{s}$.

\begin{figure}[H]
  \centering
  \includegraphics[scale=0.50]  {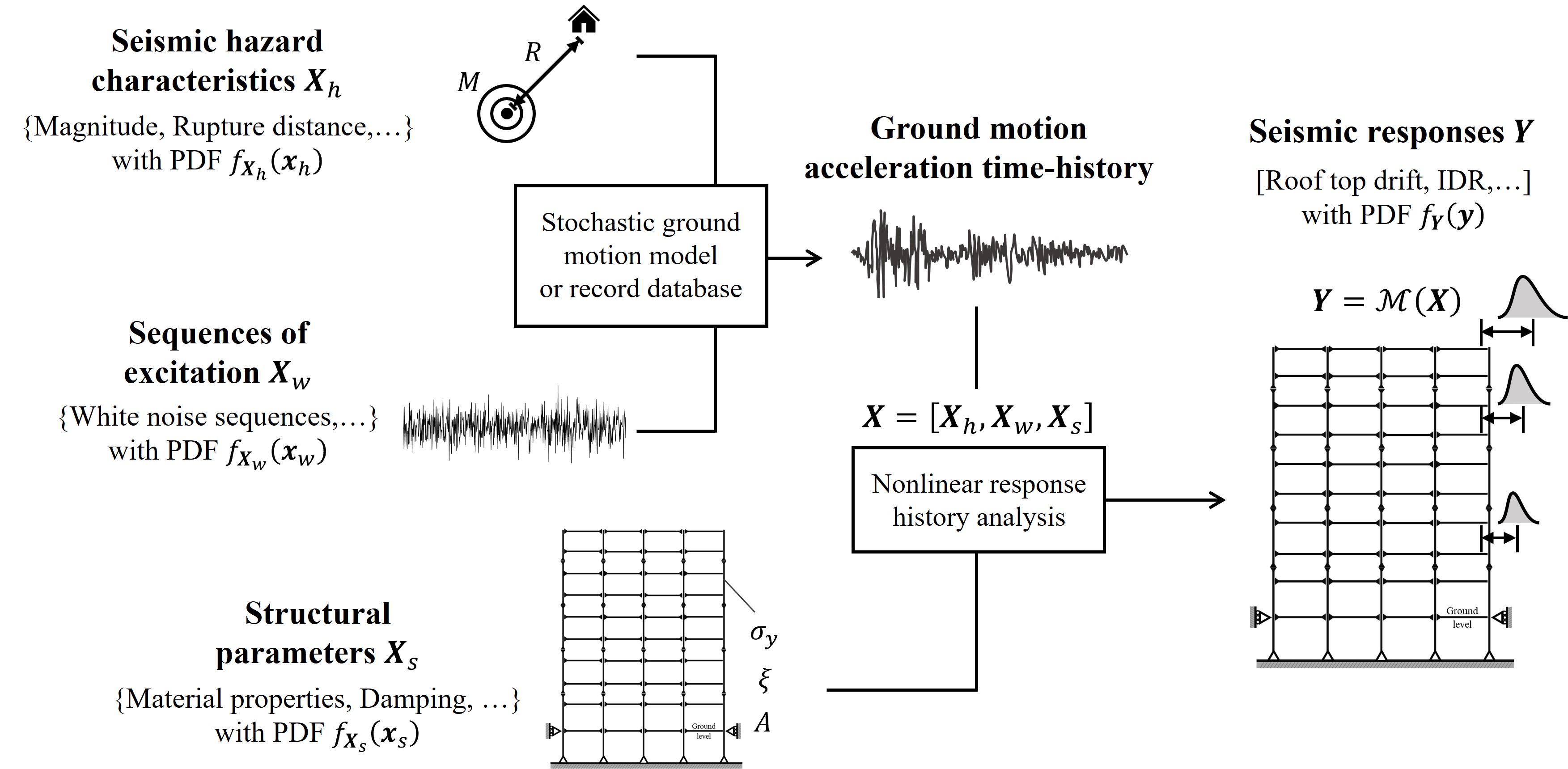}
  \caption{\textbf{Uncertainty quantification for seismic responses along with various sources of uncertainty}.}
  \label{Fig:seismicUQ}
\end{figure}

The outputs from NLRHA, $\vect{Y}\in\r^{m}$, are vectors containing Engineering Demand Parameters (EDPs) of interest, such as roof displacement and inter-story drift ratios, all of which are peak values during seismic events. The computational model underlying NLRHA is formalized by the mapping:
\begin{equation}  \label{Eq:Model}
\mathcal{M}: \vect{x}\in\rn\mapsto \vect{y}\in\r^{m} \,.
\end{equation}
Given the joint probability density function (PDF) of $\vect{X}$, the objective of UQ is to estimate the generalized moments and the joint probability distribution of $\vect{Y}$. This task is challenging due to the complexity of the computational model $\mathcal{M}$ and the high-dimensionality of $\vect{X}$. It is important to note that, regardless of the adopted ground motion models, estimating seismic responses involves addressing the high-dimensional uncertainties inherent in the ground motion processes. Figure \ref{Fig:seismicUQ} illustrates the uncertainty propagation process in seismic responses along with various sources of uncertainty.

Surrogate models are often employed to alleviate computational burdens from repeated NLRHAs and to efficiently propagate uncertainties. However, the challenge persists in approximating the end-to-end computational model in Eq.~\eqref{Eq:Model}.

\subsection{Dimensionality reduction-based surrogate modeling method (DR-SM)} \label{DRSM}

\noindent The DR-SM method \cite{kim2024dimensionality} is designed to extract a stochastic surrogate model directly from the results of the dimensionality reduction performed within the input-output space of a computational model. A key advantage of DR-SM is that it enables surrogate-based predictions for high-dimensional inputs. DR-SM uses only forward dimensionality reduction mappings, thereby circumventing the need for inverse mapping from the feature space back to the original input-output space.

\begin{figure}[H]
  \centering
  \includegraphics[scale=0.49]  {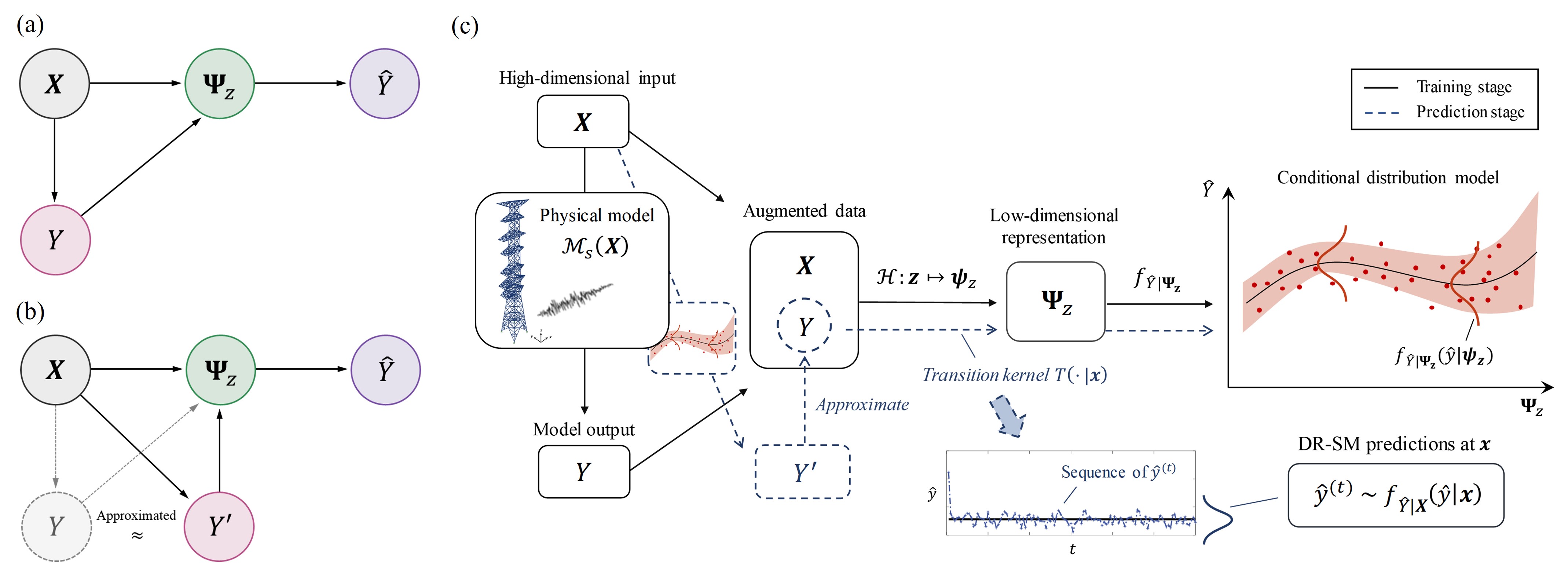}
  \caption{\textbf{Dimensionality reduction-based surrogate modeling (DR-SM) method \cite{kim2024dimensionality}: (a) the interdependency model for the training stage, and (b) the interdependency model for the prediction stage, (c) an illustration of DR-SM}. {In figure (c), the solid arrows are associated with the training stage, where the dimensionality reduction map $\mathcal{H}$ and the feature space conditional distribution $f_{\hat Y|\vect\Psi_{\vect z}}$ are obtained. The dashed arrows are associated with the prediction stage, where the trained $\mathcal{H}$ and $f_{\hat Y|\vect\Psi_{\vect z}}$ are used to generate samples from $f_{\hat Y|\vect X}(\hat y|\vect X=\vect x)$, thereby predicting $y$ given $\vect x$.}}
  \label{Fig:Illustration}
\end{figure}

The DR-SM procedure includes the following steps:
\begin{itemize}
    \item \textbf{Step 1}. Construct a training dataset of input-output pairs $\{\vect z^{(i)}\}_{i=1}^N\equiv\{(\vect{x}^{(i)},y^{(i)})\}_{i=1}^N$, where each output is directly obtained from the computational model $\mathcal{M}_s: \vect{x}\in\rn\mapsto y\in\r$, i.e., $y^{(i)}\equiv\mathcal{M}_s(\vect{x}^{(i)})$. In DR-SM, a single output $Y$ is considered. 
    
    \item \textbf{Step 2}. Implement dimensionality reduction on the input-output space, i.e., establish a mapping $\mathcal{H}:\vect z\equiv(\vect x,y)\in\mathbb{R}^{n+1}\mapsto\vect\psi_{\vect z}\in\mathbb{R}^d$. Here, $\vect\psi_{\vect z}$ represents the feature vector, and $d$ denotes the reduced dimension. Note that $\vect\psi_{\vect z}$ is contributed by both $\vect x$ and $y$. This step could accommodate various dimensionality reduction techniques, such as principal component analysis (PCA), kernel-PCA, and autoencoder.
    
    \item \textbf{Step 3}. To predict $y$ given $\vect\psi_{\vect z}$, construct a conditional distribution model $f_{\hat Y|\vect\Psi_{\vect z}}(\hat y|\vect\psi_{\vect z})$ using the dataset $\{(\vect{\psi}_{\vect z}^{(i)},y^{(i)})\}_{i=1}^N$ derived from the prior step. The heteroscedastic Gaussian process model \cite{lazaro2011variational} is employed to represent $f_{\hat Y|\vect\Psi_{\vect z}}$. Using $\mathcal{H}$ and $f_{\hat Y|\vect\Psi_{\vect z}}( \hat y|\vect\psi_{\vect z})$, the target surrogate model $f_{\hat Y|\vect X}(\hat y|\vect x)$ is formulated as:
    \begin{equation} \label{Eq:exactmod}
    f_{\hat Y|\vect X}(\hat y|\vect x)=\int f_{\hat Y|\vect\Psi_{\vect z}}(\hat y|\vect\psi_{\vect z})f_{\vect\Psi_{\vect z}|\vect XY}(\vect\psi_{\vect z}|\vect x y)f_{Y|\vect X}(y|\vect x)\,d\vect\psi_{\vect z}\,dy\,,
    \end{equation}
    where $f_{\vect\Psi_{\vect z}|\vect XY}$ corresponds to the dimensionality reduction, and $f_{Y|\vect X}$ is the computational model. Eq.~\eqref{Eq:exactmod} typically encodes an increase in uncertainty when transitioning from $f_{Y|\vect X}$ to $f_{\hat Y|\vect X}$, arising from potential errors or uncertainties introduced by the dimensionality reduction process and the feature space conditional distribution modeling. The interdependencies between $\vect X$, $Y$, $\vect\Psi_{\vect z}$, and $\hat Y$ during this step are illustrated in Figure \ref{Fig:Illustration}(a). 

    \item \textbf{Step 4}. The final step obtains a decoupled surrogate model $f_{\hat Y|\vect X}(\hat y|\vect x)$ for predicting $y$ given $\vect x$. This process is challenging as the feature vector $\vect\psi_{\vect z}$ is contributed by both $\vect x$ and $y$. DR-SM introduces a new dependency structure illustrated in Figure \ref{Fig:Illustration}(b), which leads to the following iterative equation:
    \begin{equation}  \label{Eq:approxmod}
    f_{\hat Y|\vect X}^{(t+1)}(\hat y|\vect x) = \int f_{\hat Y|\vect\Psi_{\vect z}}(\hat y|\vect\psi_{\vect z})f_{\vect\Psi_{\vect z}|\vect XY}(\vect\psi_{\vect z}|\vect x y')f_{\hat Y|\vect X}^{(t)}(y'|\vect x)\,d\vect\psi_{\vect z}\,dy' \,,
    \end{equation}
    where $f_{\hat Y|\vect X}^{(t)}$ represents an approximated surrogate model after $t$ iterations, and $y'$ is an approximation from $f_{\hat Y|\vect X}^{(t)}$. Under the presumption that $f_{\hat Y|\vect\Psi_{\vect z}}\cdot f_{\vect\Psi_{\vect z}|\vect{X} Y}$ encodes a stationary distribution, iterations of Eq.~\eqref{Eq:approxmod} lead to a fixed-point equation for the approximated surrogate model. This fixed-point equation can be interpreted as the stationarity equation for a Markov process $\{\hat{Y}^{(t)}|\vect X=\vect{x}^*\}$ with a transition kernel:   
    \begin{equation}  \label{Eq:transition}    T\left(\hat{y}^{(t)},\hat{y}^{(t+1)}|\vect{x}^*\right) = f_{\hat Y|\vect\Psi_{\vect z}}\left(\hat{y}^{(t+1)}|\vect\psi_{\vect z}\right)f_{\vect\Psi_{\vect z}|\vect{X} Y}\left(\vect\psi_{\vect z}|\vect{x}^* \hat{y}^{(t)}\right) \,. 
    \end{equation}
    This indicates that, given an unexplored input $\vect{x}^*$, a sequence of random samples generated by Eq.~\eqref{Eq:transition}, i.e., $\{\hat{y}^{(t)}\}_{t=1}^{N_t}$, can be used as the surrogate model predictions at $\vect{x^*}$, thereby constituting a stochastic surrogate model $f_{\hat Y|\vect X}(\hat y|\vect{x}^*)$. It should be noted that this sampling procedure is a special case of Markov Chain Monte Carlo (MCMC) sampling \cite{gilks1995markov}, with a transition kernel $T(\cdot|\vect{x}^*)$ derived directly from the stationarity condition.
\end{itemize}

In DR-SM, Steps 1-3 are designated as the \textit{training stage}, which can be accomplished using existing dimensionality reduction and conditional distribution modeling techniques, while Step 4 is referred to as the \textit{prediction stage}, during which the surrogate model is extracted from the results of the training stage. It should be noted that the extraction of a surrogate model in the prediction stage relies on the heuristic assumption that the transition kernel in Eq.~\eqref{Eq:transition} admits a stationary distribution. This implies that the sequence of samples generated by the transition kernel converges to a stationary distribution, provided that the dimensionality reduction is effective, i.e., the conditional distribution in the feature space provides acceptable accuracy. A detailed discussion and an idealized case study can be found in \cite{kim2024dimensionality}. The overall concept of DR-SM is illustrated in Figure \ref{Fig:Illustration}.

\section{Proposed stochastic simulator for uncertainty quantification in seismic response} \label{Proposed}

\noindent In this section, we extend the DR-SM method to (i) optimize the dimensionality reduction for seismic UQ applications, thus overcoming the limitations of traditional non-intrusive techniques, such as PCA and kernel-PCA, and (ii) predict multiple response quantities and their correlations. The resulting stochastic simulator can propagate aleatory and epistemic uncertainties related to the ground motion and structural models into multiple seismic responses, making it desirable for seismic UQ applications.

\subsection{Physics-based dimensionality reduction for ground motion uncertainties} \label{Physics-based}
\noindent The dimensionality reduction mapping $\mathcal{H}$ is crucial to the accuracy of DR-SM. Traditional dimensionality reduction techniques often fail to yield accurate surrogate models, especially when the ground motion is wideband. To address this issue, we integrate a physics-based dimensionality reduction with standard algorithms as follows:
\begin{equation}  \label{Eq:physics_DR1}
\vect{\psi}_{\vect z} = \mathcal{H}(\vect{x},\vect{y}) = \left(\mathcal{H}_{DR}\circ\mathcal{H}_p\right)(\vect{x},\vect{y}) \,,
\end{equation}
where ``$\circ$" denotes function composition. The mappings are specified as:
\begin{equation}  \label{Eq:physics_DR2}
\begin{aligned}
&\mathcal{H}_p:  \vect{x}\in\r^n\mapsto \vect{x}'\in\r^{n'} \,,\,\, n'\leq n \,,  \\
&\mathcal{H}_{DR}: (\vect{x}',\vect{y})\in\r^{n'+m}\mapsto \vect{\psi}_{\vect z}\in\r^{d} \,.
\end{aligned}
\end{equation}
$\mathcal{H}_p$ focuses on ground motion characteristics, extracting physical properties from ground motion time-histories. Subsequently, $\mathcal{H}_{DR}$ can be constructed using a conventional dimensionality reduction algorithm, which maps $(\vect{x}',\vect{y})$  to a low-dimensional representation $\vect{\psi}_{\vect z}$. Incorporating the integrated $\mathcal{H}$ into DR-SM, the method can now handle problems with high-dimensional aleatory uncertainties in the ground motion processes, whether derived from SGMM or from ground motion databases. 

In this paper, $\mathcal{H}_p$ is constructed by considering the features listed in Table \ref{Tab_Xprime}, collectively denoted by $\vect{X}'_{GM}$. These features are identified based on their significant influence on seismic responses, as reported in recent studies \cite{rezaeian2012simulation,kim2020probabilistic,sreenath2023hybrid,ding2024feature,kim2024deep}. Consequently, $\mathcal{H}_p$ maps the original input uncertainties to $\vect{X}' = [\vect{X}_{h}, \vect{X}'_{GM}, \vect{X}_{s}]$, including uncertainties from seismic events, ground motions, and structural properties. We employ PCA for $\mathcal{H}_{DR}$ due to its simplicity, although alternative methods can also be applied.

\begin{figure}[H]
  \centering
  \includegraphics[scale=0.45] {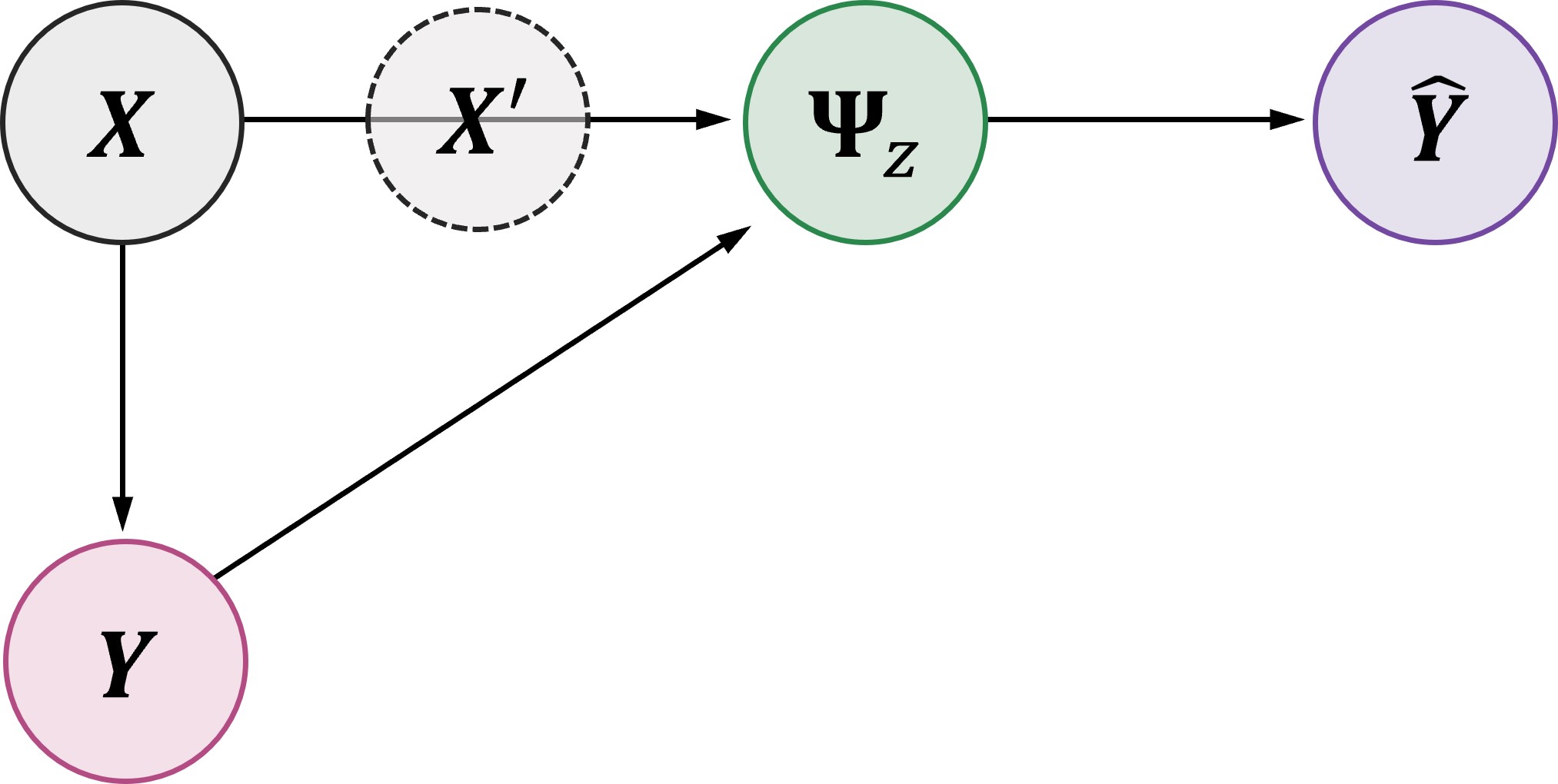}
  \caption{\textbf{Interdependency model in the proposed stochastic simulator for seismic UQ}. {Compared to the model in Figure \ref{Fig:Illustration}, the physics-based input variable $\vect{X}'$ is introduced and the outputs $\vect Y$, $\hat{\vect Y}$ are modeled as vector quantities.}}
  \label{Fig:Graph}
\end{figure}

This current formulation of $\mathcal{H}$ ensures that the most salient features of the ground motions are captured. The reduced dimension $d$ will be determined adaptively using an algorithm to be introduced in Section \ref{d_selection}. The interdependency model for the stochastic simulator is illustrated in Figure \ref{Fig:Graph}. The addition of a layer regarding $\vect{X} \mapsto \vect{X}'$ may induce additional information loss; hence, $\vect{X}'$ should be constructed with the important features of the original input.

\begin{table}[h]
  \caption{\textbf{Physics-based input variables for dimensionality reduction of ground motion uncertainties}.}   \label{Tab_Xprime}
  \centering
  \begin{tabular}{c c c c}
    \toprule
    Variables & Unit & Size & Description \\
    \midrule
    $PGA$ & $g$ & $1\times1$ & Peak ground acceleration \\
    $PGV$ & $m/s$ & $1\times1$ & Peak ground velocity \\
    $PGD$ & $m$ & $1\times1$ & Peak ground displacement \\
    $Sa(t)$ & $g$ & $1\times98$ & \begin{tabular}{@{}c@{}} Spectral acceleration of a ground motion with 5\% damping\\from period of $0.01$ to $10$ seconds \end{tabular} \\
    $I_A$ & $m/s$ & $1\times1$ & \begin{tabular}{@{}c@{}} Total arias intensity, i.e., time integral of squared ground motion\\acceleration $a(t)$ over the duration $T_{d}$, i.e., $I_A = \frac{\pi}{2g}\int_{0}^{T_{d}} a(t)^2\,dt$ \end{tabular} \\
    $mean(I_A)$ &$m/s$ & $1\times1$ & Mean of $I_a$ \\
    $median(I_A)$ & $m/s$ & $1\times1$ & Median of $I_a$ \\
    $t_{mean}$ & $s$ & $1\times1$ & Time when mean of $I_A$ has occurred  \\
    $t_{median}$ & $s$ & $1\times1$ & Time when median of $I_A$ has occurred \\
    $D_{5-75}$ & $s$ & $1\times1$ & Duration between when 5\% and 75\% of $I_A$ has occurred \\
    $D_{5-95}$ & $s$ & $1\times1$ & Duration between when 5\% and 95\% of $I_A$ has occurred \\
    $t_{45}$ & $s$ & $1\times1$ & Time when 45\% of $I_A$ has occurred \\
    \bottomrule
  \end{tabular} \\
  \raggedright{\hspace{0.7cm} {\footnotesize * $g$ is the gravitational acceleration.}} \\
\end{table}

\subsection{Conditional distribution model for multivariate outputs} \label{Condi_GMM}

\noindent The DR-SM method is primarily designed for single-output problems. To enhance its applicability to multivariate output scenarios, which is crucial for seismic UQ, we adopt a multivariate conditional distribution model that can predict the vector response $\vect{y}$ given the features $\vect{\psi}_{\vect{z}}$.

From the training dataset $\{(\vect{x}^{(i)},\vect{y}^{(i)})\}_{i=1}^N$, where $\vect{y}^{(i)}=\mathcal{M}(\vect{x}^{(i)})$, the feature-output set $\{(\vect{\psi}_{\vect z}^{(i)},\vect{y}^{(i)})\}_{i=1}^N$ can be obtained by applying $\mathcal{H}$ as described in Eq.~\eqref{Eq:physics_DR1}. In this multi-dimensional space $(\vect{\Psi}_{\vect z},\vect{Y}) \in \mathbb{R}^{d+m}$, we first construct a Gaussian mixture model (GMM) \cite{sung2004gaussian,hu2019probability} to approximate the joint PDF of features and responses, $f_{\vect{\Psi}_{\vect z} \hat{\vect{Y}}}(\vect{\psi}_{\vect{z}},\hat{\vect{y}})$, expressed as:

\begin{equation}  \label{Eq:GMM_joint}
f_{\vect\Psi_{\vect z} \hat{\vect Y}}(\vect\psi_{\vect z},\hat{\vect y}) =  \sum_{i=1}^{q} \pi^i f_N(\vect\psi_{\vect z},\hat{\vect y};\vect{\mu}^i,\vect{\Sigma}^i)  \,,
\end{equation}

\noindent where $f_N$ denotes the Gaussian PDF, $q$ is the number of mixture components, and $\pi^i$, $\vect{\mu}^i$, $\vect{\Sigma}^i$ are the mixture weight, mean vector, and covariance matrix for each component, respectively, as determined by the expectation-maximization algorithm. The mean vector and covariance matrix for each component are structured as follows:

\begin{equation}  \label{Eq:GMM_muCOV}
\vect{\mu}^i = \left[ \vect{\mu}_{\vect\Psi_{\vect z}}^i,\, \vect{\mu}_{\hat{\vect Y}}^i \right], \,\,\, \vect{\Sigma}^i = 
\begin{bmatrix}
\vect{\Sigma}_{\vect\Psi_{\vect z}\vect\Psi_{\vect z}}^i & \vect{\Sigma}_{\vect\Psi_{\vect z}\hat{\vect Y}}^i  \\
\vect{\Sigma}_{\hat{\vect Y}\vect\Psi_{\vect z}}^i  & \vect{\Sigma}_{\hat{\vect Y}\hat{\vect Y}}^i 
\end{bmatrix}
\,.
\end{equation}

Given $\vect\Psi_{\vect z} = \vect\psi_{\vect z}$, the conditional PDF $f_{\hat{\vect Y}|\vect\Psi_{\vect z}}(\hat{\vect y}|\vect\psi_{\vect z})$ is derived from the joint distribution as follows:

\begin{equation}  \label{Eq:GMM_condi}
f_{\hat{\vect Y}|\vect\Psi_{\vect z}}(\hat{\vect y}|\vect\psi_{\vect z}) = \sum_{i=1}^{q} \pi^i_{\hat{\vect Y}|\vect\Psi_{\vect z}} f_N(\hat{\vect y};\vect{\mu}^i_{\hat{\vect Y}|\vect\Psi_{\vect z}},\vect{\Sigma}^i_{\hat{\vect Y}|\vect\Psi_{\vect z}})  \,,
\end{equation}

\noindent where $\pi^i_{\hat{\vect Y}|\vect\Psi_{\vect z}}$, $\vect{\mu}^i_{\hat{\vect Y}|\vect\Psi_{\vect z}}$, $\vect{\Sigma}^i_{\hat{\vect Y}|\vect\Psi_{\vect z}}$ represent the conditional mixture weight, mean vector, and covariance matrix, respectively. These parameters are analytically computed using the parameters of the GMM model:

\begin{equation}  \label{Eq:Condi_weight}
\pi^i_{\hat{\vect Y}|\vect\Psi_{\vect z}} = \frac{\pi^i f_N(\vect\psi_{\vect z},\vect{\mu}_{\vect\Psi_{\vect z}}^i,\vect{\Sigma}_{\vect\Psi_{\vect z}\vect\Psi_{\vect z}}^i)} {\sum_{k=1}^{q} \pi^k f_N(\vect\psi_{\vect z},\vect{\mu}_{\vect\Psi_{\vect z}}^k,\vect{\Sigma}_{\vect\Psi_{\vect z}\vect\Psi_{\vect z}}^k)}  \,,
\end{equation}

\noindent and

\begin{equation}  \label{Eq:Condi_muCOV}
\begin{aligned}
&\vect{\mu}^i_{\hat{\vect Y}|\vect\Psi_{\vect z}} = \vect{\mu}_{\hat{\vect Y}}^i + \vect{\Sigma}_{\hat{\vect Y}\vect\Psi_{\vect z}}^i \vect{\Sigma}_{\vect\Psi_{\vect z}\vect\Psi_{\vect z}}^{i^{-1}} \left( \vect\psi_{\vect z} - \vect{\mu}_{\vect\Psi_{\vect z}}^i \right)  \,,  \\
&\vect{\Sigma}^i_{\hat{\vect Y}|\vect\Psi_{\vect z}} = \vect{\Sigma}_{\hat{\vect Y}\hat{\vect Y}}^i - \vect{\Sigma}_{\hat{\vect Y}\vect\Psi_{\vect z}}^i \vect{\Sigma}_{\vect\Psi_{\vect z}\vect\Psi_{\vect z}}^{i^{-1}} \vect{\Sigma}_{\vect\Psi_{\vect z}\hat{\vect Y}}^{i}   \,.
\end{aligned}
\end{equation}
To summarize, Eqs.~\eqref{Eq:GMM_condi}, \eqref{Eq:Condi_weight}, and \eqref{Eq:Condi_muCOV} enable DR-SM to predict multivariate seismic response quantities.

\subsection{Determination of the reduced dimension} \label{d_selection}
\noindent Identifying the ideal dimension of $\vect{\Psi_z}$ is crucial for the accuracy of the proposed stochastic simulator. Building upon previous studies on dimensionality reduction-based UQ methods \cite{kim2024adaptive,kim2024dimensionality}, we introduce Algorithm \ref{alg:dimension} to estimate the optimal reduced dimension $d$. The main idea is to find the most parsimonious representation, i.e., the smallest $d$, that achieves an acceptable prediction accuracy. For the $k$-th output, $Y_k$, we consider the following mean squared error:

\begin{equation}  \label{Eq:error_d}
\varepsilon_k^{d} = \sqrt{\frac{1}{N}\sum_{i=1}^{N}\left(y_k^{(i)}-\mu_{\hat{Y}_k|\vect\Psi_{\vect z}}\left({\vect\psi}_{\vect{z}}^{(i)}; d \right)\right)^2} \,,
\end{equation}

\noindent where $\mu_{\hat{Y}_k|\vect\Psi_{\vect z}}$ denotes the mean prediction for $Y_k$ given $\vect{\Psi}_{\vect{z}}$, obtained from the mean of the conditional distribution in Eq.~\eqref{Eq:GMM_condi}, expressed as:

\begin{equation}  \label{Eq:GMM_mean}
\vect{\mu}_{\hat{\vect{Y}}|\vect\Psi_{\vect z}}(\vect\psi_{\vect z}) = \mathbb{E}[\hat{\vect{Y}}|\vect\psi_{\vect z}] = \sum_{i=1}^{q} \pi^i_{\hat{\vect Y}|\vect\Psi_{\vect z}} \vect{\mu}^i_{\hat{\vect Y}|\vect\Psi_{\vect z}}  \,,
\end{equation}

\noindent where the parameters $\pi^i_{\hat{\vect{Y}}|\vect{\Psi}_{\vect{z}}}$ and $\vect{\mu}^i_{\hat{\vect{Y}}|\vect{\Psi}_{\vect{z}}}$ are expressed by Eq.~\eqref{Eq:Condi_muCOV}.

We adopt an average error measure, expressed as $\varepsilon^{d} = \frac{1}{m} \sum_{k=1}^{m} \varepsilon_k^{d}$, to quantify the impact of the dimension $d$ on the accuracy of the feature space conditional distribution model. We start with $d=1$ and iteratively increase $d$ until $\varepsilon^{d}$ falls below a specified threshold $\varepsilon^{d}_t$, thereby balancing accuracy and model complexity in seismic response predictions. This procedure is summarized in Algorithm \ref{alg:dimension}.

\begin{breakablealgorithm}
\label{alg:dimension}
\caption{Adaptive procedure to determine the optimal reduced dimension $d^{*}$.}
\begin{description}
Given a set of training data $\{{\vect{x}'}^{(i)} ,\vect{y}^{(i)}\}_{i=1}^{N}$, where ${\vect{x}'}^{(i)}=\mathcal{H}_p(\vect{x}^{(i)})$:
\\
$d\leftarrow{0}$; $\varepsilon^{d}\leftarrow\infty$;\\
\textbf{While} $\varepsilon^{d}>\varepsilon^{d}_t$, \textbf{do}
\\
\,\,\,\,\,\,\,$d\leftarrow{d+1}$; 
\\
\,\,\,\,\,\,\,Identify the $d$-dimensional feature mapping;
\\
\,\,\,\,\,\,\,Compute the feature space mean predictions $\vect{\mu}_{\hat{\vect{Y}}|\vect\Psi_{\vect z}}({\vect\psi}_{\vect{z}}^{(i)}; d)$;
\\
\,\,\,\,\,\,\,Compute the average error $\varepsilon^{d}$;
\\
\textbf{End}
\\
$d^*\leftarrow{d}$; 
\end{description}
\end{breakablealgorithm}

\subsection{Algorithm of the stochastic simulator for seismic UQ applications} \label{Algorithm}

\begin{figure}[h]
  \centering
  \includegraphics[scale=0.60] {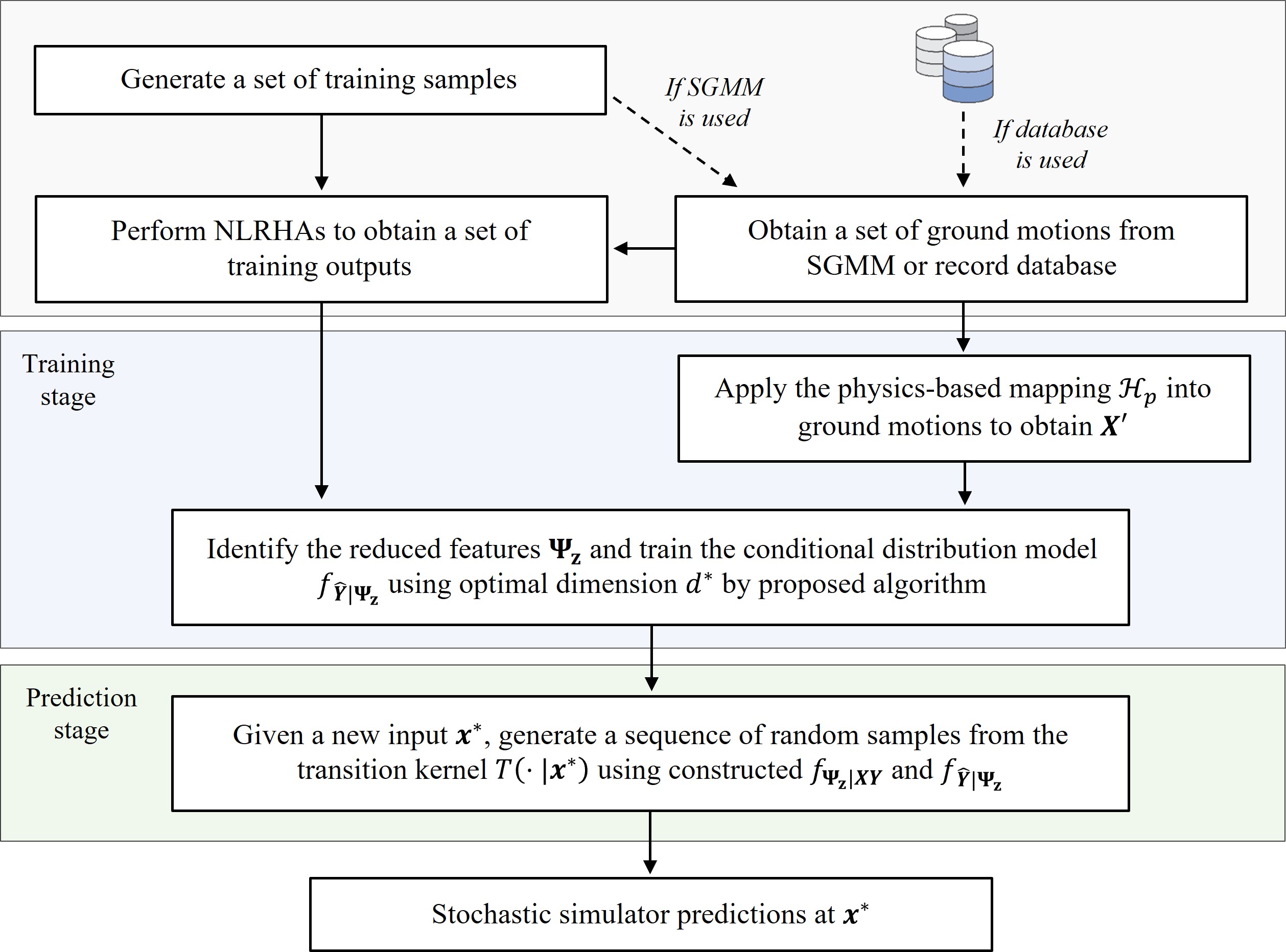}
  \caption{\textbf{Flowchart of the proposed stochastic simulator approach}.}
  \label{Fig:algorithm}
\end{figure}

\noindent Provided with the integrated dimensionality reduction mapping $\mathcal{H}$ and the multivariate conditional distribution model $f_{\hat{\vect{Y}}|\vect{\Psi}_{\vect{z}}}(\hat{\vect{y}}|\vect{\psi}_{\vect{z}})$, we summarize the DR-SM approach below and in Figure \ref{Fig:algorithm}.

\begin{description}
\item [1. Construct a training dataset]
~\
\begin{itemize}
\item Generate training sets for seismic hazard and structural parameters, i.e., $\mathcal{X}_{h} = \{\vect{x}_{h}^{(i)}\}_{i=1}^{N}$ and $\mathcal{X}_{s} = \{\vect{x}_{s}^{(i)}\}_{i=1}^{N}$. The selection of the training sample size involves a trade-off between computational efficiency and prediction accuracy, typically influenced by the dimensionality of the inputs and outputs, as well as the complexity of the structural system. We observed that a sample size of $N \in (400, 800)$ is sufficient for the problems we have investigated.
\item Obtain a set of ground motions:
\begin{itemize}
\item For artificial (simulated) ground motions: Utilize SGMM to simulate $N$ ground motions based on samples from $\mathcal{X}_h$. A set of samples $\mathcal{X}_{w} = \{\vect{x}_{w}^{(i)}\}_{i=1}^{N}$ is used to simulate stochastic sequences for each training point.
\item For real (recorded) ground motions: Aggregate $N$ ground motion records from a database. In this case, the ground motions are regarded as random realizations of $\vect{X}_h$ and $\vect{X}_w$.
\end{itemize}
\item Perform NLRHA on each sample to obtain the seismic response set $\mathcal{Y} = \{\mathcal{M}(\mathcal{X})\}$, where $\mathcal{X}=(\mathcal{X}_{h},\mathcal{X}_{w},\mathcal{X}_{s})$.
\end{itemize}

\item [2. Identify the reduced features and train conditional distribution model]
~\
\begin{itemize}
\item Apply the physics-based dimensionality reduction $\mathcal{H}_p$ to the simulated/recorded ground motions to obtain the dataset $\mathcal{Z}'_\mathcal{D}=\{({\vect{x}'}^{(i)},\vect{y}^{(i)})\}_{i=1}^N$.
\item Using $\mathcal{Z}'_\mathcal{D}$ and Algorithm \ref{alg:dimension}, determine the optimal reduced dimension $d^*$ and train the multivariate conditional distribution model. This process establishes the feature mapping $f_{\vect\Psi_{\vect z}|\vect{X}\vect{Y}}$ and the feature space conditional distribution  $f_{\hat{\vect{Y}}|\vect\Psi_{\vect z}}$.
\end{itemize}

\item [3. Extract a stochastic surrogate model for seismic responses]
~\
\begin{itemize}
\item For a new input $\vect{x}^*=(\vect{x}_h^*,\vect{x}_w^*,\vect{x}_s^*)$, generate a sequence of random samples $\{\hat{\vect{y}}^{(t)}\}_{t=1}^{N_t}$ using the transition kernel described in Eq.~\eqref{Eq:transition}, replacing $\hat{y}$ with $\hat{\vect{y}}$. The sequence length is set to $N_t=1300$, with the first 300 samples designated as burn-in—the predictions are thus obtained from $t=301,...,1300$. The starting point $\hat{\vect y}^{(1)}$ is set as the mean of the training set. 

\item Use these samples as stochastic simulator predictions at $\vect{x}^*$: This provides the statistical quantities of the outputs, including the mean prediction $\vect{\mu}_{\hat{\vect{Y}}}(\vect{x}^*) = \mathbb{E}[\hat{\vect{Y}}|\vect{x}^*]$, and the covariance matrix $\vect{\Sigma}^2_{\hat{\vect{Y}}\hat{\vect{Y}}}(\vect{x}^*) = \mathbb{V}\text{ar}[\hat{\vect{Y}}|\vect{x}^*]$, whose variance vector is $\vect{\sigma}^2_{\hat{\vect{Y}}}(\vect{x}^*)$.

\end{itemize}
\end{description}

This algorithm is designed to be adaptable, allowing for the replacement of the feature set listed in Table \ref{Tab_Xprime} with alternatives, extending its applicability beyond seismic UQ problems. The essential criterion for the method is that the selected features should reflect domain knowledge, so that efficiency and interpretability can be balanced.

\section{Numerical investigations} \label{Examples}
The performance of the proposed stochastic simulator is demonstrated through applications to three different structural systems: a three-story steel frame, a nine-story building, and a high-rise transmission tower. Each case is tested under both synthetically generated ground motions through the SGMM and recorded ground motions from a database, illustrating the simulator's versatility and relevance in earthquake engineering practice. The training samples are generated using Latin Hypercube sampling with sample decorrelation. The reference solutions are obtained through direct Monte Carlo simulations (MCS), using $20,000$ NLRHAs, which achieve a coefficient of variation of $7\%$ at a probability level of $0.01$.

For all examples, we adopt the relative mean squared errors (RMSE) to examine the accuracy of the proposed method for each output $Y_k$:
\begin{equation} \label{Eq:error_model}
\varepsilon^{\eta} =\frac{\mathbb{E}\left[\left(\mu_{\hat{Y}_k}(\vect{X}) - Y_k)\right)^{2}\right]}{\mathbb{V}\text{ar}\left[Y_k\right]}\,,
\end{equation}
\noindent where $\mu_{\hat{Y}_k}$ denotes the mean prediction for $Y_k$.

\subsection{Application to three-story steel frame structure} \label{3SAC}
\begin{figure}[H]
  \centering
  \includegraphics[scale=0.60] {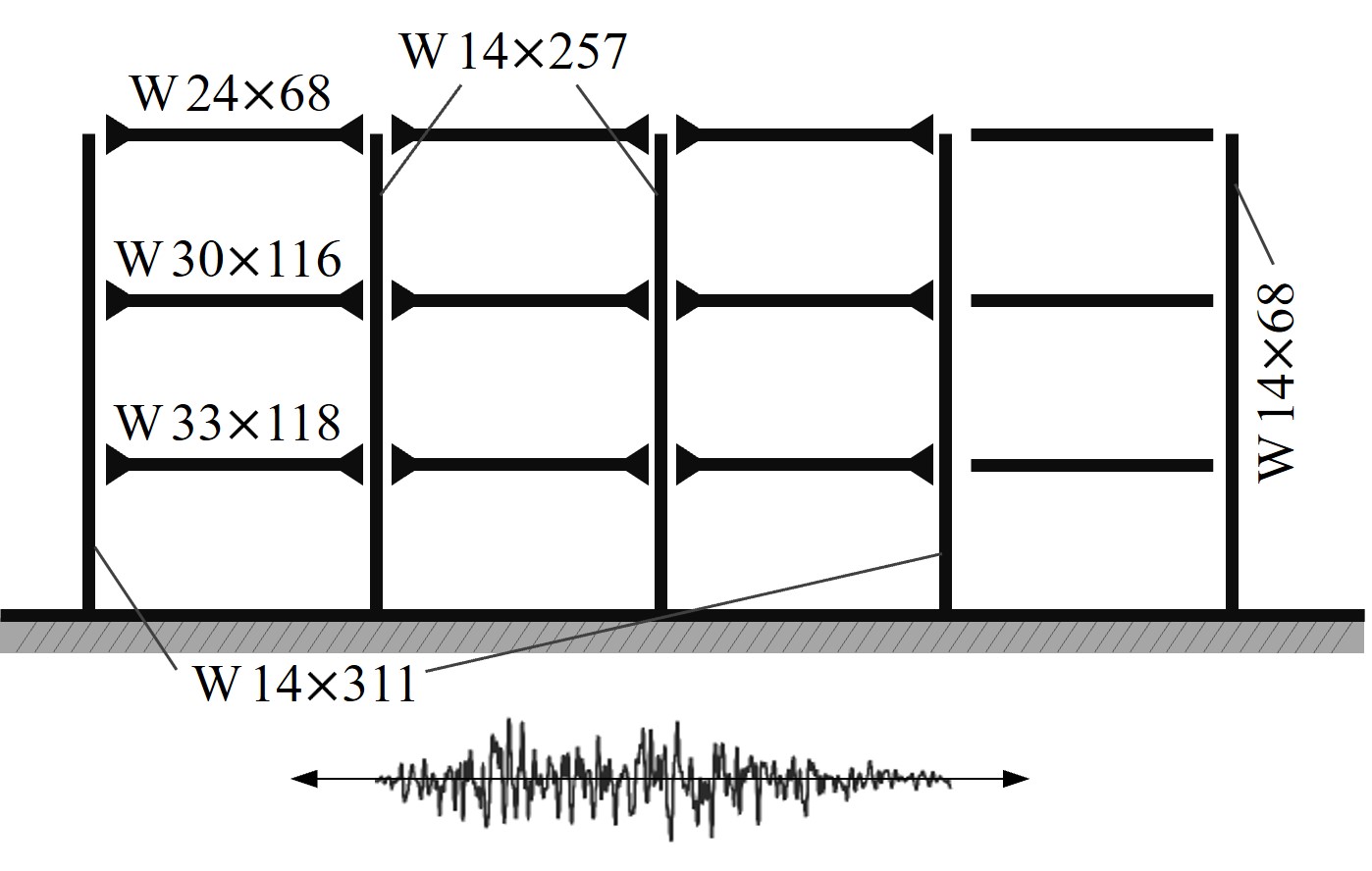}
  \caption{\textbf{A three-story steel MRF structure}.}
  \label{Fig_3SAC}
\end{figure}

Consider a three-story steel moment-resisting frame (MRF) structure \cite{ohtori2004benchmark} subjected to seismic loads, designed as a benchmark structure of the SAC joint venture project (Figure \ref{Fig_3SAC}). The structure consists of three-story, four-bay frames with each story’s height and bay’s span being 3.96 m and 9.15 m, respectively. A rigid diaphragm is assumed in the structural model, which is created by the OpenSees software to perform the NLRHA including the P-delta effect. The seismic responses include the peak inter-story drift ratios (IDR) for each story, represented as $\vect{Y}=\{IDR_1,IDR_2,IDR_3\}\in\r^3$.

\subsubsection{Case 1: Artificial ground motions generated by SGMM} \label{3SAC_SGMM}

\noindent We first consider a scenario in which the SGMM is employed to generate artificial ground motions. For the SGMM, a point source model \cite{boore2003simulation} is adopted, which characterizes the temporal envelope and radiation spectrum of stochastic excitation. This characterization depends on two parameters: earthquake magnitude $M$ and rupture distance $R_{rup}$, which collectively form the excitation model parameter set $\vect{X}_h$. For the rupture distance $R_{rup}$, a lognormal distribution is assumed, whereas the magnitude $M$ is described by the Gutenberg-Richter model truncated between $[M_{min}, M_{max}] = [6,8]$, leading to the PDF $f_M(M)=\lambda_M \exp{(-\lambda_M M)}/\left[\exp{(-\lambda_M M_{min})} - \exp{(-\lambda_M M_{max})} \right]$ \cite{kyprioti2021kriging,yi2024stochastic}. The seismic rate parameter is set to $\lambda_M=0.9\ln{10}$. Additionally, the vector $\vect{X}_s$ encompasses the white Gaussian noise samples utilized to generate stochastic sequences. For the structural model, the damping ratio $\xi$ and material properties—modulus of elasticity $E$, yield stress $\sigma_y$, and strain hardening ratio $\varepsilon_h$ of steels used in beam and column members—are considered as random variables that form $\vect{X}_s$. The distributions and parameters for the comprehensive set of random variables are summarized in Table \ref{Tab_3SACrvs}.

\begin{table}[H]
  \caption{\textbf{Distribution models and parameters of the random variables for the three-story steel MRF structure (Case 1)}.}
  \label{Tab_3SACrvs}
  \centering
  \begin{tabular}{c c c c c}
    \toprule
    Categories & Random variables & Parameter1 & Parameter2 & Distribution \\
    \midrule
    $\vect{X}_{h}$ & Magnitude, $M \,$ & 6 & 8 & Gutenberg–Richter \\
    $\vect{X}_{h}$ & Rupture distance, $R_{rup} \, (km)$ & 20 & 2 & Lognormal \\
    $\vect{X}_{w}$ & Gaussian noise sequences, $\phi_i \,$ & 0 & 1 & Gaussian \\
    $\vect{X}_{s}$ & Damping ratio, $\xi \, (\%)$ & 3 & 0.6 & Lognormal \\
    $\vect{X}_{s}$ & Modulus of elasticity, $E \, $(Mpa) & 200000 & 10000 & Lognormal \\
    $\vect{X}_{s}$ & Yield stress for beam, $\sigma_y^b \,$(Mpa) & 248 & 24.8 & Lognormal \\
    $\vect{X}_{s}$ & Yield stress for column, $\sigma_y^c \,$(Mpa) & 345 & 34.5 & Lognormal \\
    $\vect{X}_{s}$ & Straining hardening ratio for beam, $\varepsilon_h^b \,$ & 0.01 & 0.002 & Lognormal \\
    $\vect{X}_{s}$ & Straining hardening ratio for column, $\varepsilon_h^c \,$ & 0.01 & 0.002 & Lognormal \\
    \bottomrule
  \end{tabular}  \\
  \raggedright{\hspace{0.1cm} {\footnotesize * Parameter1 and Parameter2 respectively represent the lower and upper bounds for the Gutenberg–Richter model, while these parameters represent the mean and standard deviation for other probabilistic models.}} \\
  \raggedright{\hspace{0.1cm} {\footnotesize * The Gaussian noise $\vect X_w$ has a dimension $n_w \geq 1,000$.}} \\   
\end{table}

\begin{figure}[H]
  \centering
  \includegraphics[scale=0.45] {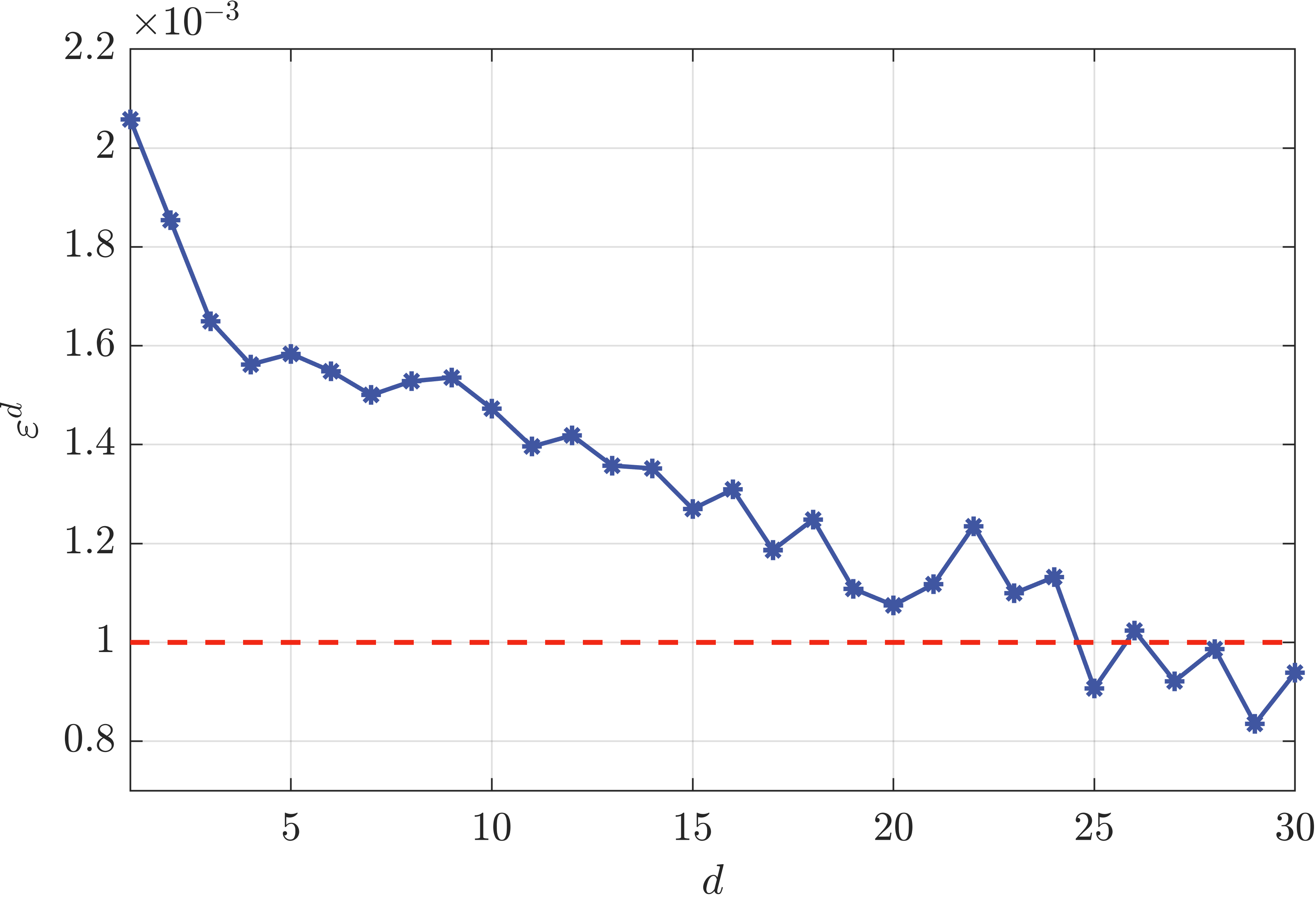}
  \caption{\textbf{Error $\varepsilon^d$ as a function of reduced dimension for the three-story steel MRF example}. {This figure is a byproduct of Algorithm 1. The red dashed line denotes the threshold level of $\varepsilon^d_t=0.001$, suggesting $d=25$ is sufficient.}}
  \label{Fig_3SAC_d}
\end{figure}

Following the procedure described in Section \ref{Proposed}, the proposed approach is implemented with a training set of $500$ samples. The physics-based dimensionality reduction $\mathcal{H}_{p}$ maps the generated stochastic ground motions into the features listed in Table \ref{Tab_Xprime}, yielding $\vect{X}'$ with a dimension of $n'=117(=2+109+6)$. Subsequently, PCA maps $(\vect{x}',\vect{y}) \in \mathbb{R}^{120(=117+3)}$ into a $d$-dimensional feature space. The reduced dimension is determined with a threshold $\varepsilon^d_t=0.001$, as described by Algorithm \ref{alg:dimension}. Figure \ref{Fig_3SAC_d} shows the variation in the error $\varepsilon^d$, which suggests that a reduced dimension of $d=25$ is needed to achieve a small prediction error.

\begin{figure}[H]
  \centering
  \includegraphics[scale=0.62] {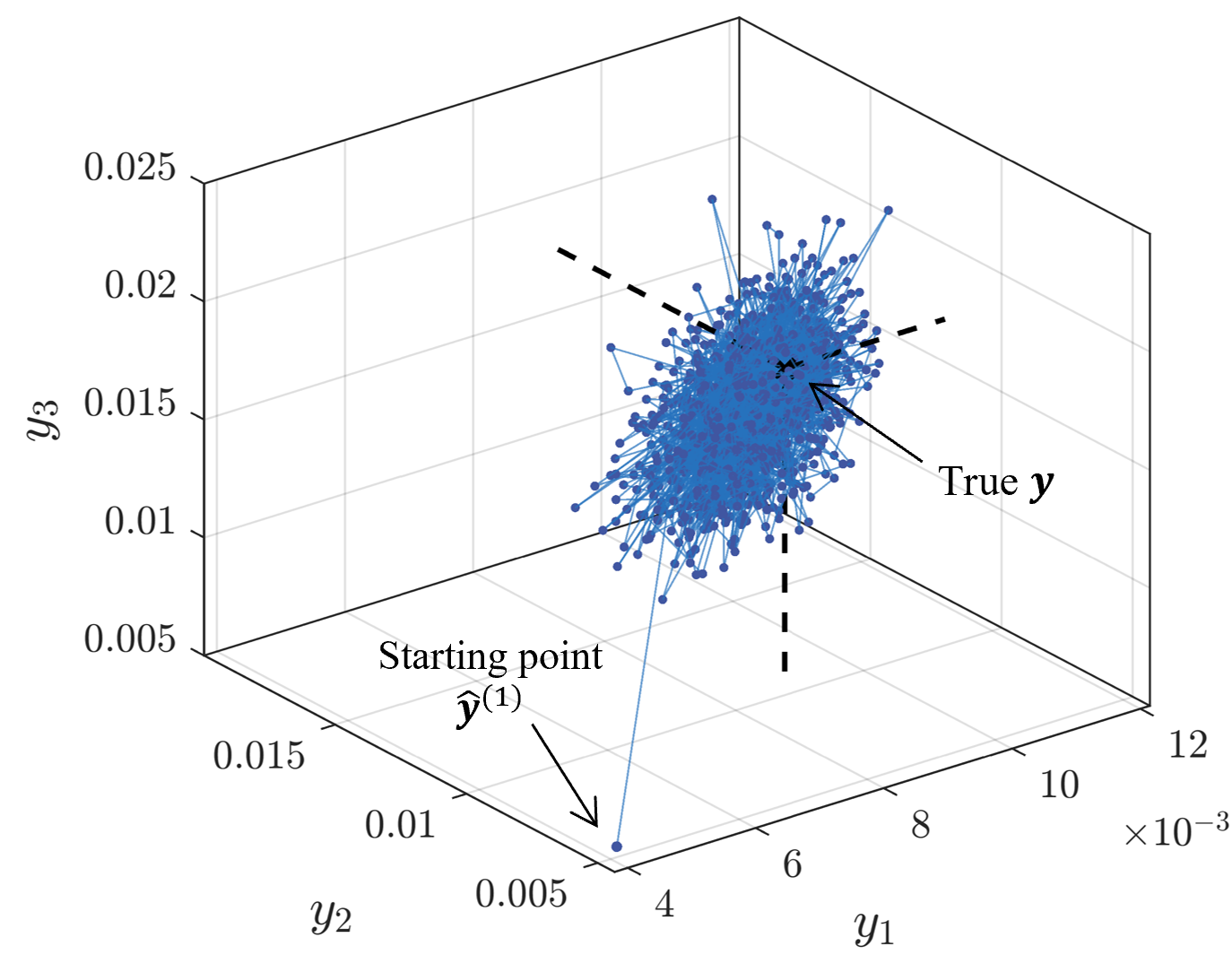}
  \caption{\textbf{Trajectories of $\hat{\vect y}$ obtained from the stochastic simulator for the three-story steel MRF example}. {The plot shows the random samples $\hat{\vect y}^{(t)}$ generated by the transition kernel $T(\hat{\vect y}^{(t)},\hat{\vect y}^{(t+1)}| \vect{x}^*)$ for a given test sample $\vect{x}^*$. The reference response values are \{0.00920, 0.01107, 0.01763\}, while the corresponding mean predictions are \{0.00891, 0.01164, 0.01598\}.}}
  \label{Fig_3SAC_chain}
\end{figure}

Figure \ref{Fig_3SAC_chain} illustrates the trajectories of random sequences $\hat{\vect y}^{(t)}, t=1,...,1300$ for a selected input $\vect{x}^*$, generated from the transition kernel $T(\hat{\vect y}^{(t)}, \hat{\vect y}^{(t+1)}| \vect{x}^*)$ as detailed in Step 3 of the algorithm in Section \ref{Algorithm}. The predictions are obtained from the generated samples, with relative errors of  $3.15\%$, $5.14\%$, and $9.36\%$ for each EDP, calculated as $|y_k - \mu_{\hat{Y}_k}(\vect{x}^*)|/{y_k}, k=1,2.3$. Note that the prediction uncertainties, i.e., the variability in the sample trajectories, stem from imperfections in the dimensionality reduction and conditional distribution modeling. The results show that the sequences converge closely to the true responses, thereby confirming the predictive accuracy of the stochastic simulator. Notably, the prediction uncertainty is directly quantified from the samples without the need for additional statistical methods, such as bootstrap resampling.

\begin{figure}[H]
  \centering
  \includegraphics[scale=0.41] {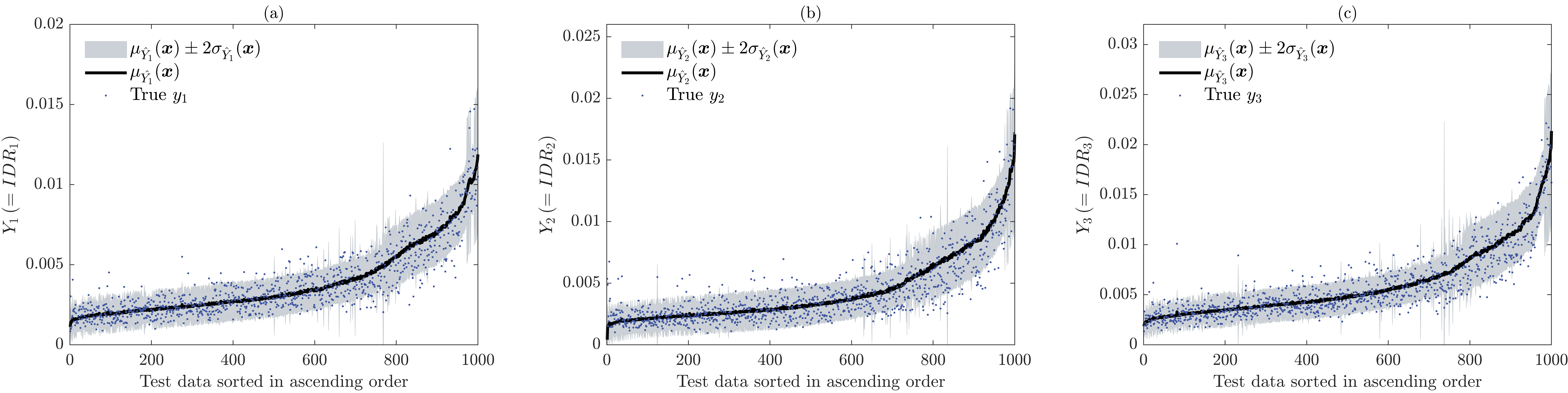}
  \caption{\textbf{Scatter plots of the true responses against the stochastic simulator predictions for the three-story steel MRF example (Case 1): (a) $IDR_1$, (b) $IDR_2$, and (c) $IDR_3$}. {The proposed simulator is trained using $500$ samples. For each plot, mean predictions and their uncertainty intervals are represented by black lines and gray shaded areas, respectively, while the true responses are denoted by blue circles. RMSE values are calculated as $0.1708$, $0.2106$, and $0.1589$ for each IDR.}}
  \label{Fig_3SAC_surrogate_SGMM}
\end{figure}

\begin{figure}[H]
  \centering
  \includegraphics[scale=0.42] {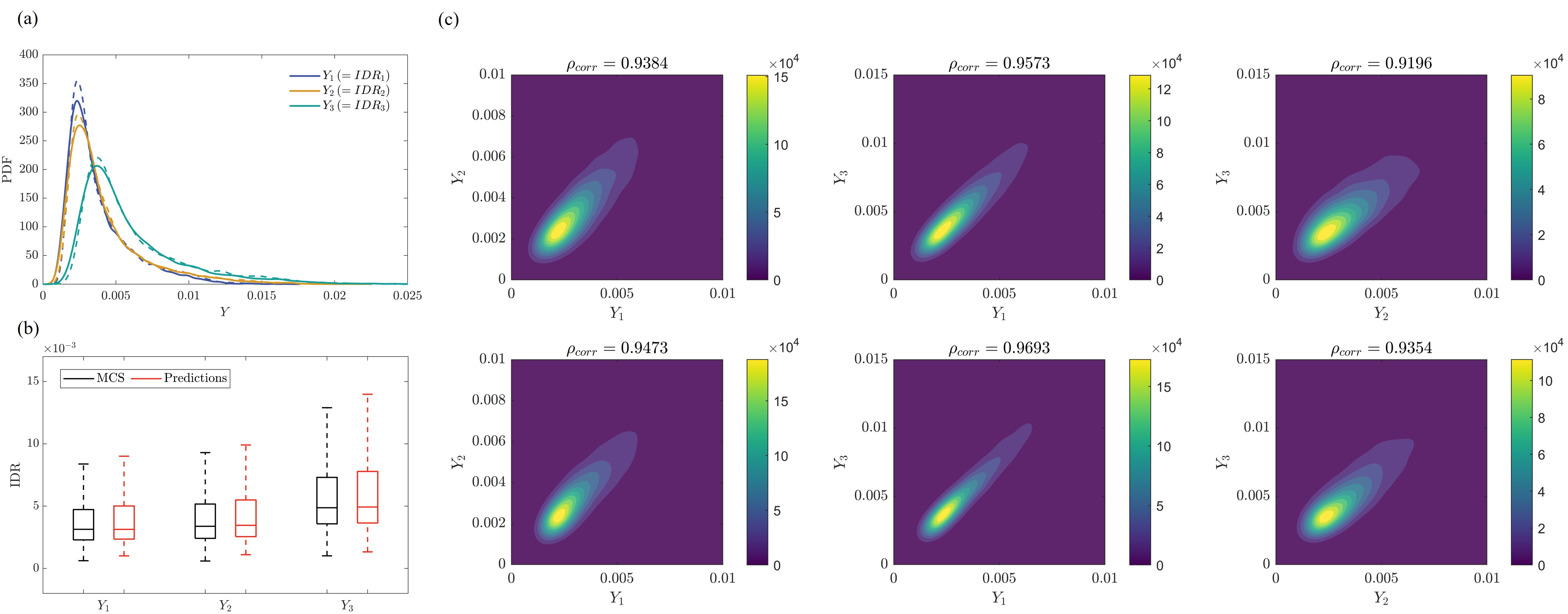}
  \caption{\textbf{Uncertainty quantification in seismic responses for the three-story steel MRF example (Case 1): (a) marginal PDFs, (b) median and interquartile ranges, and (c) joint PDFs and response correlations}. {In figure (a), references obtained by MCS and predictions by the proposed method are represented by solid and dashed lines, respectively. In figure (c), the first and second rows show the MCS references and the predictions by the proposed method.}}
  \label{Fig_3SAC_UQ}
\end{figure}

Figure \ref{Fig_3SAC_surrogate_SGMM} compares the stochastic simulator predictions against the true seismic responses. The mean predictions and standard deviation intervals are depicted by black solid lines and gray shaded areas, respectively. The true responses are denoted by blue circles, rearranged in ascending order of the predicted means. The results demonstrate that the proposed simulator can effectively capture the global trend of the true responses without overfitting. It is also observed that most true responses fall within the predicted standard deviation intervals. It should be noted that the proposed approach requires only one simulator to estimate all response quantities. Figure \ref{Fig_3SAC_UQ} presents the results of the stochastic simulator-based UQ analysis, displaying marginal PDFs, median and 25\%/75\% quantiles through boxplots, and joint PDFs with Pearson correlation coefficients. These results, compared against those from MCS, confirm the accuracy of the stochastic simulator in capturing response distributions and interdependencies between responses. It is noteworthy that the proposed approach does not assume a typical probabilistic distribution of response quantities.

\begin{figure}[H]
  \centering
  \includegraphics[scale=0.48] {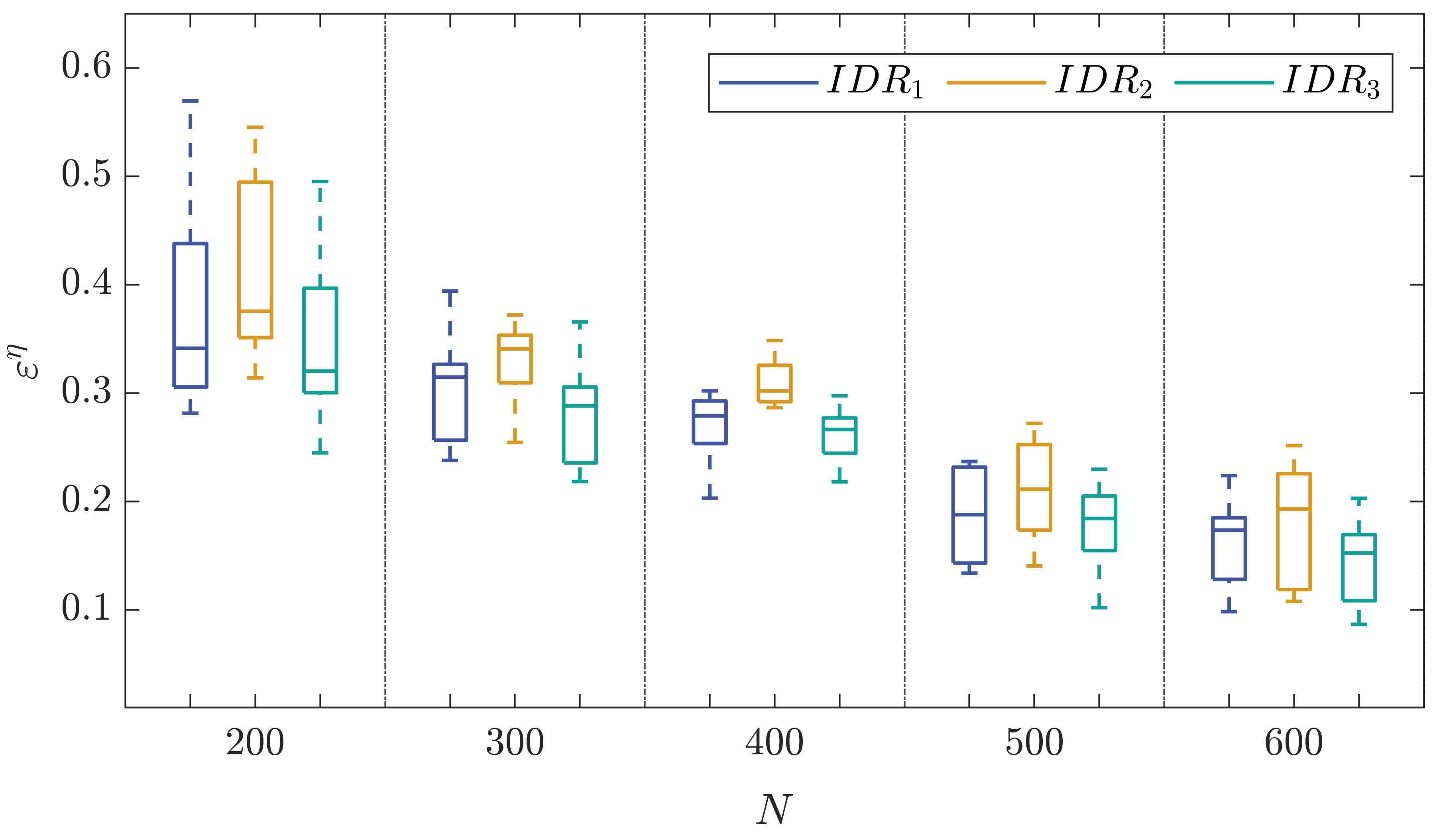}
  \caption{\textbf{Relative mean squared errors of each response at different training sample sizes for the three-story steel MRF example (Case 1)}. {Each box plot is obtained using 10 independent runs of the proposed approach.}}
  \label{Fig_3SAC_errorbox_SGMM}
\end{figure}

The RMSE for each response, defined by Eq.~\eqref{Eq:error_model}, at varying training sample sizes is illustrated in the boxplot of Figure \ref{Fig_3SAC_errorbox_SGMM}. Each plot summarizes the RMSE from ten replications. It is observed that the prediction accuracy increases with the number of training samples, demonstrating that the proposed method can efficiently quantify seismic response uncertainty using a limited computational budget.

A parametric study is conducted to further examine the effect of the convergence threshold $\varepsilon^{d}_t$ used in Algorithm \ref{alg:dimension}. Table \ref{Tab_3SACrMSE} presents the RMSEs corresponding to different threshold values: $\{1.0, 1.25, 1.5, 2.0\} \times 10^{-3}$. The reduced dimensionalities determined by these thresholds are $2$, $10$, $17$, and $25$, respectively. The table shows the mean RMSE values from $10$ replications, with standard deviations provided in parentheses. It is observed that the highest accuracy is achieved with the most conservative threshold, while larger thresholds result in reduced accuracy and greater variability in the predictions. Since overly tight convergence tolerances can lead to overfitting in both the dimensionality reduction and feature space conditional distribution modeling, a threshold of $\varepsilon^{d}_t=10^{-3}$ is recommended in this study.

\begin{table}[H]
  \caption{\textbf{Relative mean squared errors of each response for the three-story steel MRF example with different threshold values (Case 1).}}
  \label{Tab_3SACrMSE}
  \centering
  \begin{tabular}{ c c c c c }
    \toprule
    Threshold, $\varepsilon^{d}_t$ $(10^{-3})$ & Reduced dimensionality, $d$ & \multicolumn{3}{c}{RMSE, $\varepsilon^{\eta}$} \\
    \cmidrule(lr){3-5}
     & & $IDR_1$ & $IDR_2$ & $IDR_3$ \\
    \midrule
    $2.0$ & $2$ & 0.249 & 0.275 & 0.258 \\
     &  & (0.185) & (0.210) & (0.171) \\
    $1.5$ & $10$ & 0.212 & 0.261 & 0.243 \\
     &  & (0.090) & (0.115) & (0.097) \\
    $1.25$ & $17$ & 0.190 & 0.220 & 0.191 \\
     &  & (0.048) & (0.065) & (0.059) \\
    $1.0$ & $25$ & 0.189 & 0.213 & 0.179 \\
     &  & (0.041) & (0.045) & (0.039) \\
    \bottomrule
  \end{tabular}
\end{table}

\subsubsection{Case 2: Recorded ground motions from a database}  \label{3SAC_record}

\noindent Next, we examine a scenario wherein the recorded ground motions are utilized. For this analysis, a total of 2086 ground  acceleration time-histories and their earthquake characteristics, specifically magnitude and rupture distance, are collected from the PEER Next Generation Attenuation (NGA)-West2 database \cite{ancheta2014nga} based on the following criteria: $6 \leq M \leq 8$, and $10$ km $\leq R_{rup} \leq 50$ km. These datasets serve as the realizations of random vectors $\vect{X}_h$ and $\vect{X}_w$ from an unknown model. The structural model parameters listed in Table \ref{Tab_3SACrvs} are considered again as random variables within $\vect{X}_s$.

\begin{figure}[H]
  \centering
  \includegraphics[scale=0.40] {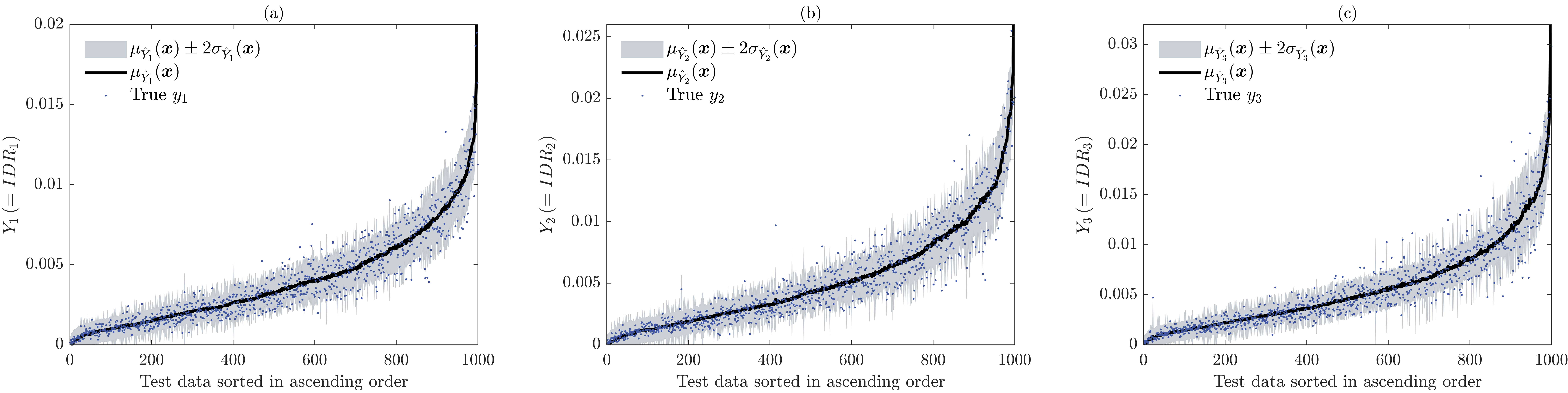}
  \caption{\textbf{Scatter plots of the true responses against the stochastic simulator predictions for the three-story steel MRF example (Case 2): (a) $IDR_1$, (b) $IDR_2$, and (c) $IDR_3$}. {The proposed simulator is trained using $600$ samples. For each plot, mean predictions and their uncertainty intervals are represented by black lines and gray shaded areas, respectively, while the true responses are denoted by blue circles. RMSE values are calculated as $0.1361$, $0.1562$, and $0.1605$ for each IDR.}}
  \label{Fig_3SAC_surrogate_record}
\end{figure}

\begin{figure}[H]
  \centering
  \includegraphics[scale=0.48] {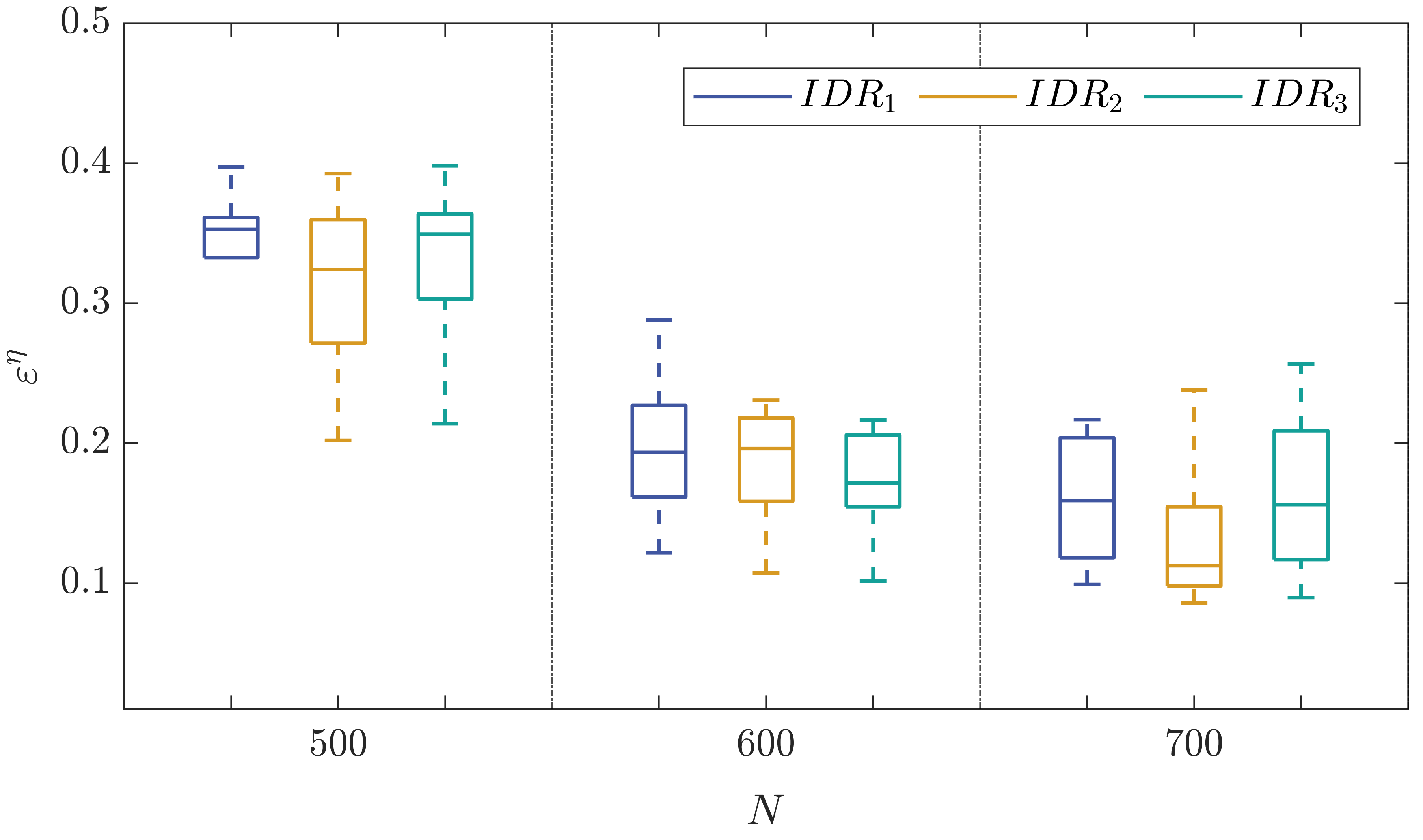}
  \caption{\textbf{Relative mean squared errors of each response at different training sample sizes for the three-story steel MRF example (Case 2)}. {Each box plot is obtained using 10 independent runs of the proposed approach.}}
  \label{Fig_3SAC_errorbox_record}
\end{figure}

Similarly to case 1, the proposed simulator is initiated with a training set of 600 samples. The physics-based dimensionality reduction is directly applied to the ground motion records. The optimal reduced dimension is determined to be $d=27$, using a threshold $\varepsilon^d_t=0.001$. Figure \ref{Fig_3SAC_surrogate_record} presents a comparison between the predicted means and prediction intervals against true responses, using 1,000 test samples from the ground motion database. Figure \ref{Fig_3SAC_errorbox_record} shows the RMSEs $\varepsilon^{\eta}$ across various training sample sizes. The results demonstrate that the proposed stochastic simulator is capable of accurately replicating the global trend of the seismic responses under real recorded ground motions.


\subsection{Application to a nine-story steel building structure} \label{9SAC}
\begin{figure}[H]
  \centering
  \includegraphics[scale=0.37] {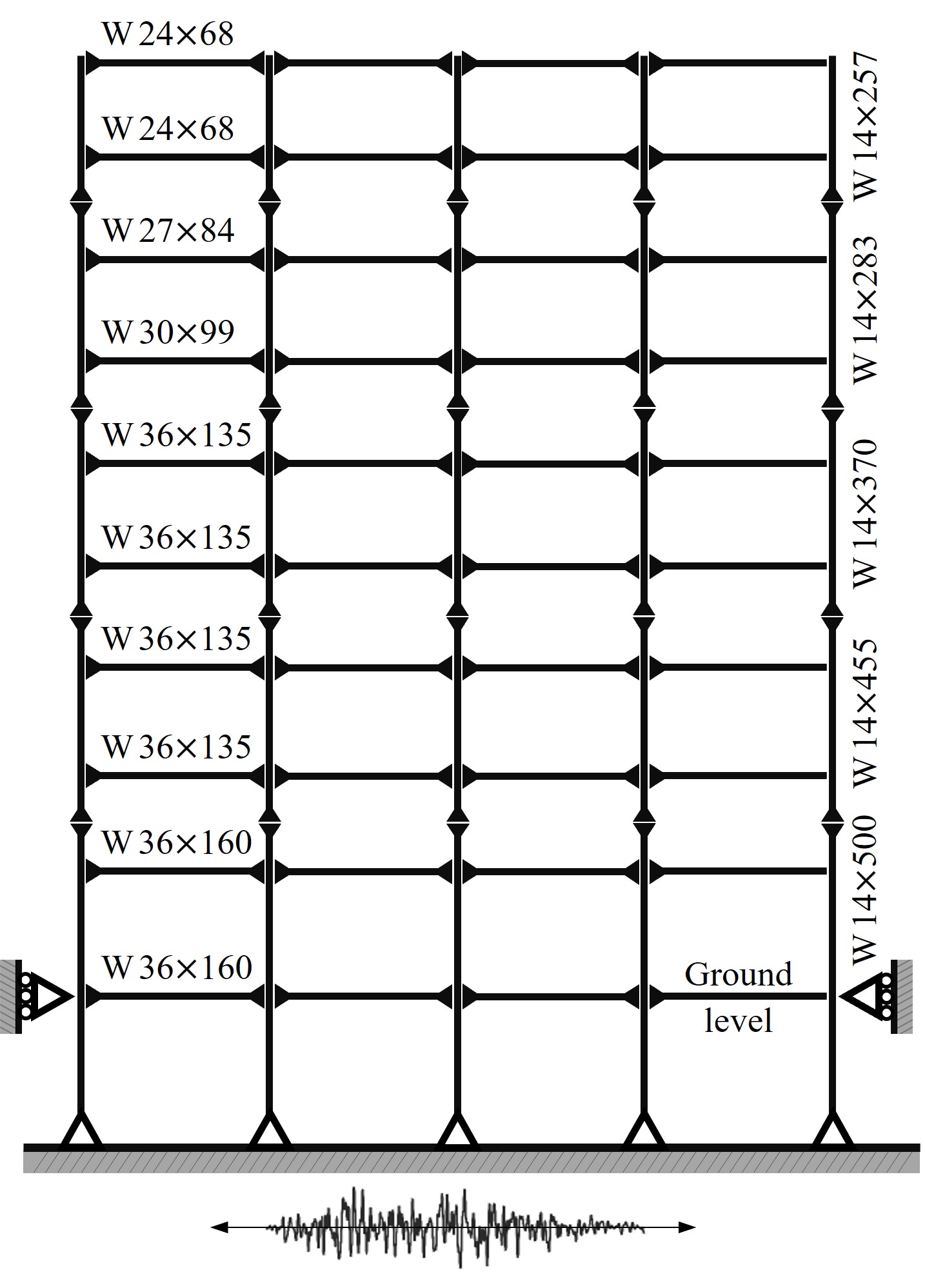}
  \caption{\textbf{A nine-story steel building structure}.}
  \label{Fig_9SAC}
\end{figure}

\noindent This application examines a nine-story steel building structure \cite{ohtori2004benchmark}, shown in Figure \ref{Fig_9SAC}, to investigate the performance of the proposed method in a structural system characterized by higher mode effects. The bay width and elevation are 45.73 m and 37.19 m, respectively. To effectively carry bending and uplift forces from seismic excitation, the column joints of the splice story are located on the first, third, fifth, and seventh levels, each elevated 1.83 m above the beam's centerline. The assumed foundational supports, in the form of concrete walls and surrounding soil, act to restrain the ground level of the structural system. The seismic responses include the peak interstory drift ratio (IDR) and peak story displacement (SD) for each story, i.e., $\vect{Y} = \{IDR_1, \dots, IDR_9, SD_1, \dots, SD_9\} \in \mathbb{R}^{18}$. In the structural model, the damping ratio and the material properties of beam and column elements are treated as random variables, with their distributions specified in Table \ref{Tab_3SACrvs}.

\subsubsection{Case 1: Artificial ground motions generated by SGMM}

\noindent Employing the point source SGMM with uncertain excitation model parameters $M$ and $R_{rup}$ in Section \ref{3SAC_SGMM}, the proposed simulator is calibrated with a dataset of 600 training samples. The reduced dimension is identified as $d=33$, obtained with a threshold $\varepsilon^d_t=0.001$.

\begin{figure}[H]
  \centering
  \includegraphics[scale=0.40] {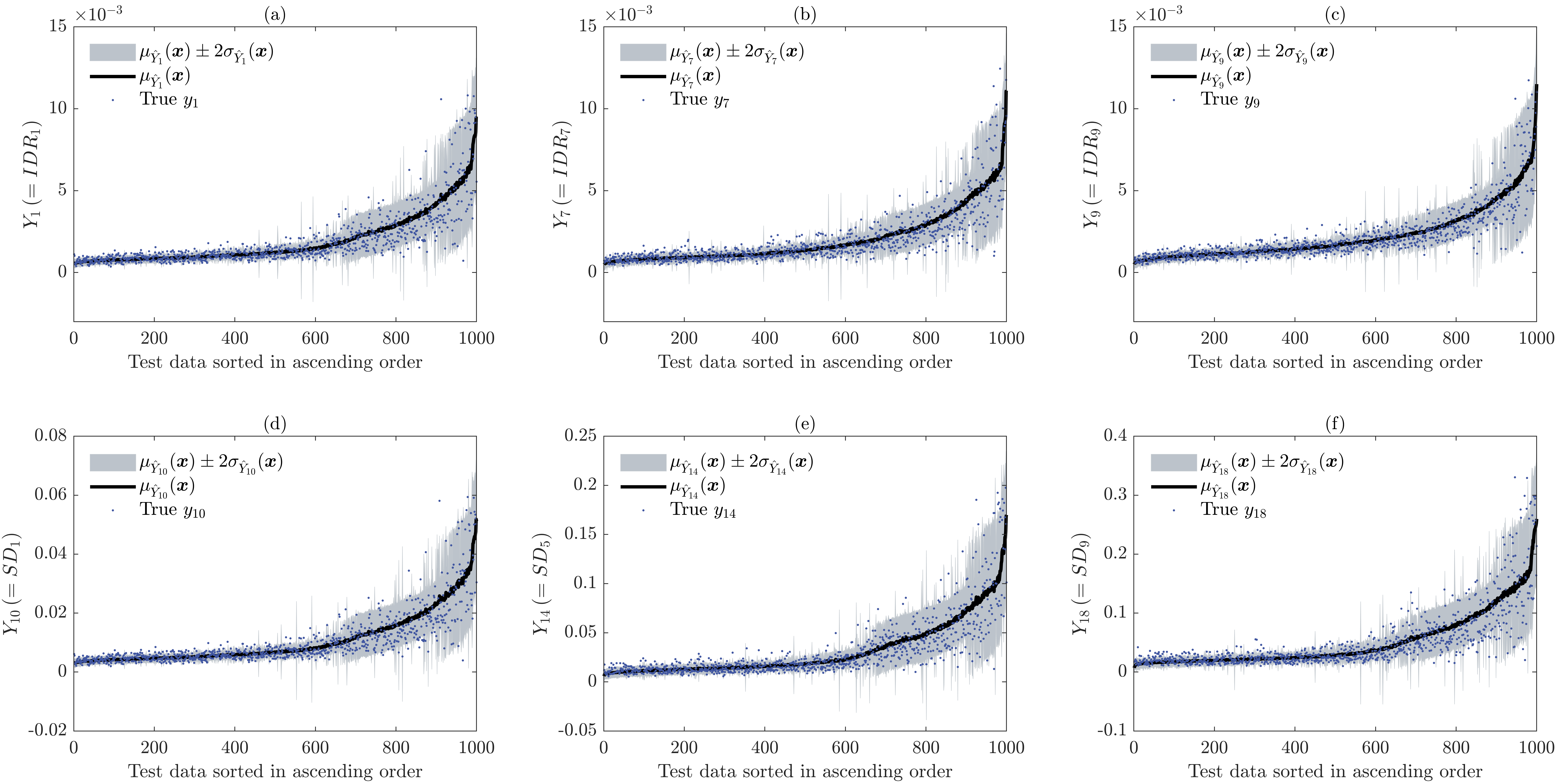}
  \caption{\textbf{Scatter plots of the true responses against the stochastic simulator predictions for the nine-story steel building example (Case 1): (a) $IDR_1$, (b) $IDR_7$, (c) $IDR_9$, (d) $SD_1$, (e) $SD_5$ and (c) $SD_9$}. {This plot shows six responses from $\vect{Y}=\{Y_1,...,Y_{18}\}$. The proposed simulator is trained with $700$ samples. For each plot, mean predictions and their uncertainty intervals are represented by black lines and gray shaded areas, respectively, while the true responses are denoted by blue circles. RMSE values are calculated as $0.2337$, $0.2222$, $0.1793$, $0.2317$, $0.2429$ and $0.2570$ for each response.}}
  \label{Fig_9SAC_surrogate_SGMM}
\end{figure}

\begin{figure}[H]
  \centering
  \includegraphics[scale=0.45] {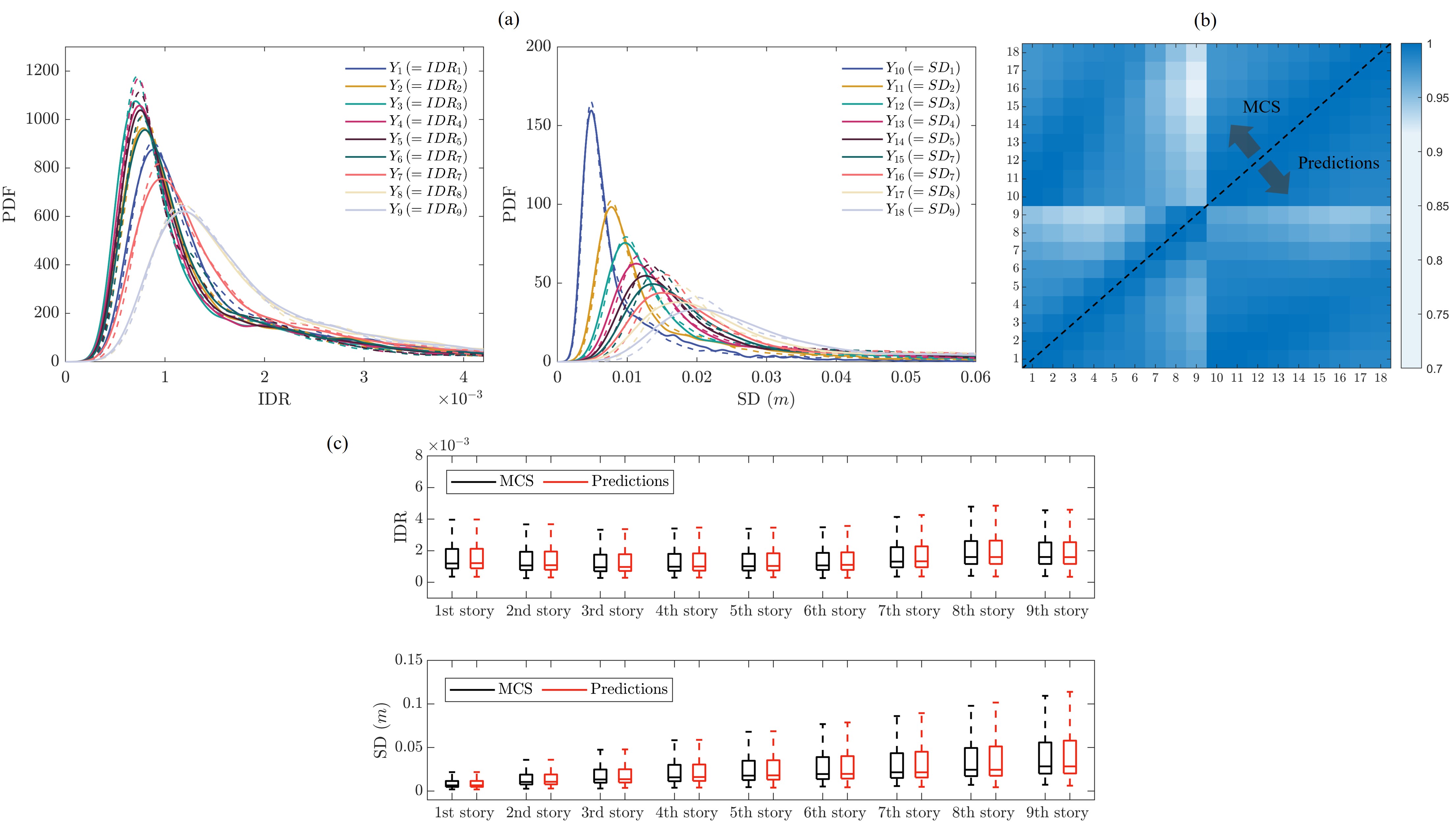}
  \caption{\textbf{Uncertainty quantification in seismic responses for the nine-story steel building example (Case 1): (a) marginal PDFs, (b) correlation coefficient matrix for responses, and (c) median and interquartile ranges}. {In figure (a), references obtained by MCS and predictions by the proposed method are represented by solid and dashed lines, respectively. In figure (b), the axis labels denote the respective response variables from $\vect{Y}=\{Y_1,...,Y_{18}\}$.}}
  \label{Fig_9SAC_UQ0}
\end{figure}

\begin{figure}[H]
  \centering
  \includegraphics[scale=0.48] {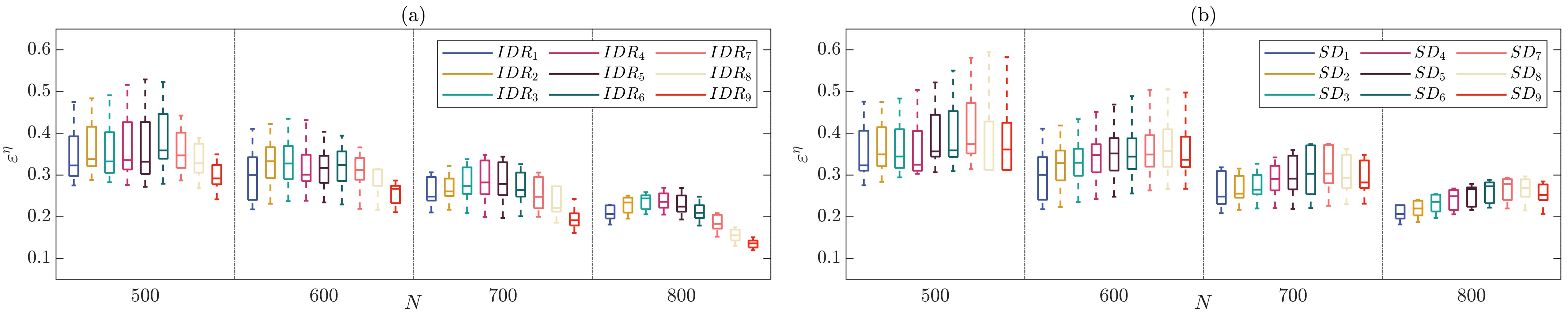}
  \caption{\textbf{Relative mean squared errors of each response at different training sample sizes for the nine-story steel building example (Case 1)}. {Each box plot is obtained using 10 independent runs of the proposed approach.}}
  \label{Fig_9SAC_errorbox_SGMM}
\end{figure}

Figure \ref{Fig_9SAC_surrogate_SGMM} presents the scatter plots of the predicted seismic responses and prediction intervals, compared with the true values. Compared to the previous example, we observed large prediction intervals in high peak responses due to the nonlinearity and insufficient training data. The estimated marginal distributions, Pearson correlation coefficient matrix, and median with interquartile ranges are detailed in Figure \ref{Fig_9SAC_UQ0}. Additionally, Figure \ref{Fig_9SAC_errorbox_SGMM} compares the RMSEs across varying training sample sizes. The results confirm the accuracy and efficiency of the proposed method. It is noted that the proposed simulator can handle different scales of seismic responses simultaneously, including both ratios and absolute displacement values.


\subsubsection{Case 2: Recorded ground motions from a database}

\begin{figure}[H]
  \centering
  \includegraphics[scale=0.43] {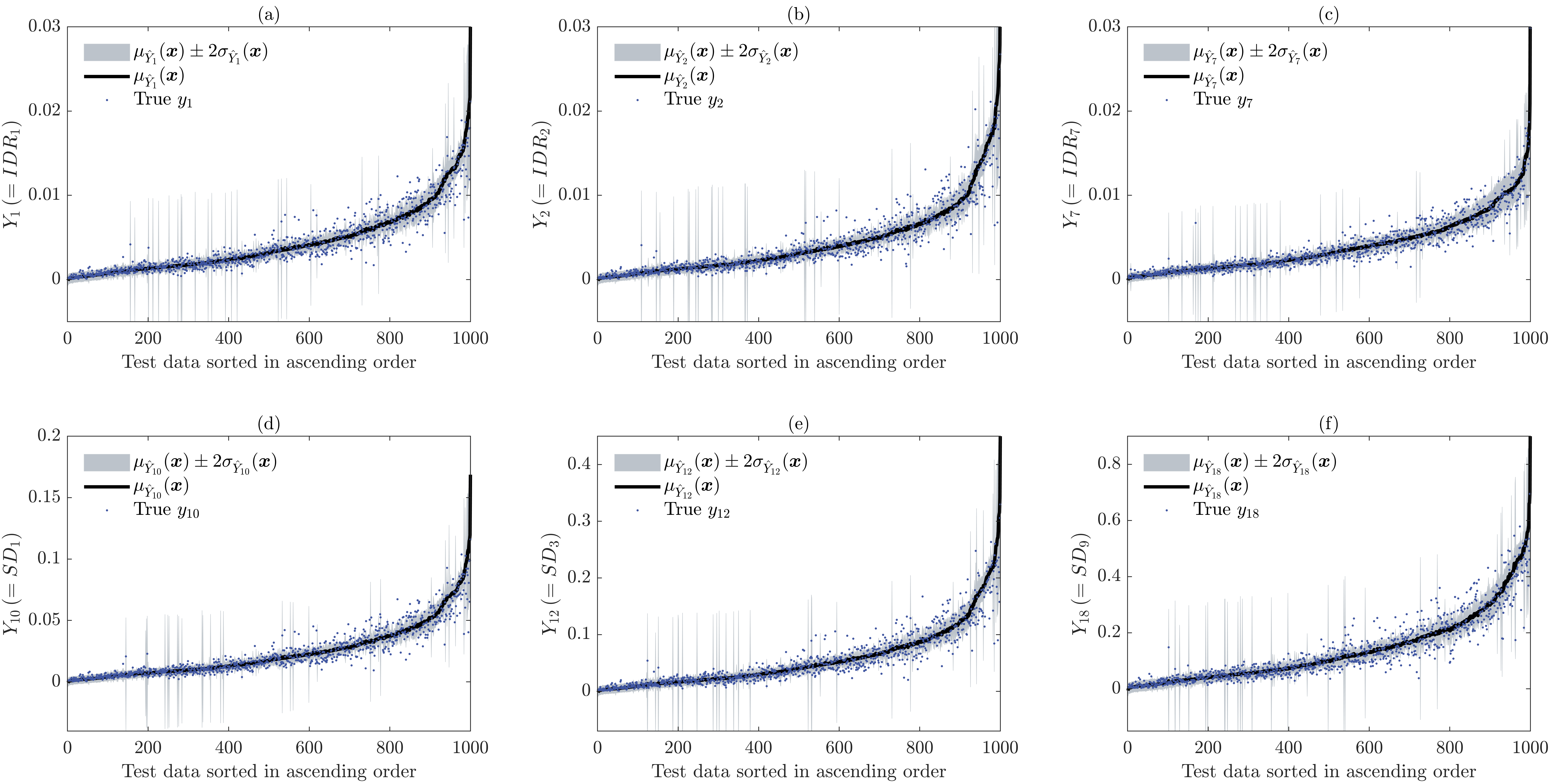}
  \caption{\textbf{Scatter plots of the true responses against the stochastic simulator predictions for the nine-story steel building example (Case 2): (a) $IDR_1$, (b) $IDR_2$, (c) $IDR_7$, (d) $SD_1$, (e) $SD_3$ and (c) $SD_9$}. {This plot shows six responses from $\vect{Y}=\{Y_1,...,Y_{18}\}$. The proposed simulator is trained using $700$ samples. For each plot, mean predictions and their uncertainty intervals are represented by black lines and gray shaded areas, respectively, while the true responses are denoted by blue circles. RMSE values are calculated as $0.1246$, $0.1487$, $0.0886$, $0.1247$, $0.1418$ and $0.1110$ for each response.}}
  \label{Fig_9SAC_surrogate_record}
\end{figure}

\begin{figure}[H]
  \centering
  \includegraphics[scale=0.48] {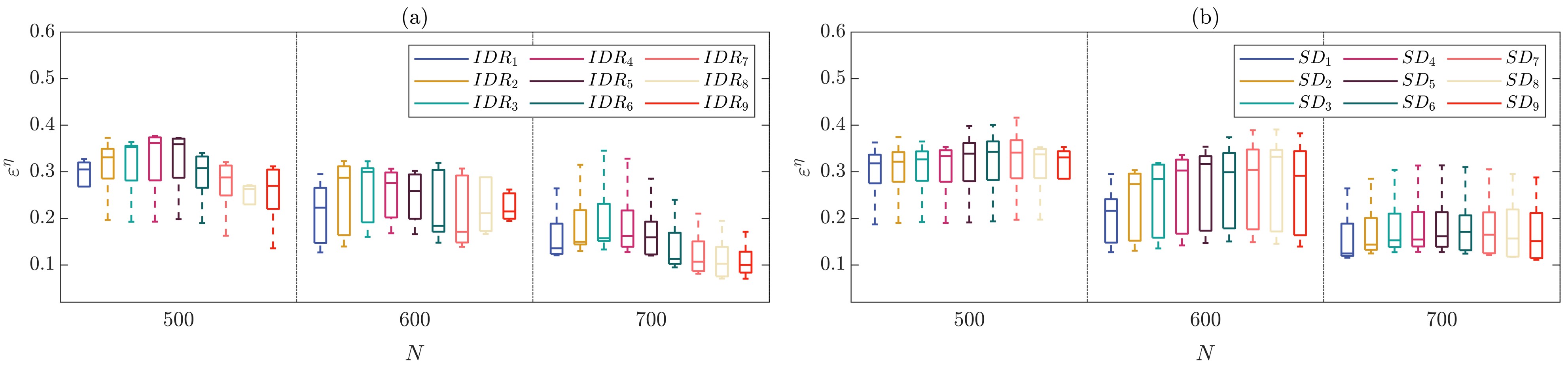}
  \caption{\textbf{Relative mean squared errors of each response at different training sample sizes for the nine-story steel building example (Case 2)}. {Each box plot is obtained using 10 independent runs of the proposed approach.}}
  \label{Fig_9SAC_errorbox_record}
\end{figure}

\noindent The proposed simulator is applied to a dataset comprising recorded ground motions used in Section \ref{3SAC_record}. The efficiency of the simulator is tested using 700 training data and 1,000 test data. Similar to the previous application, Figure \ref{Fig_9SAC_surrogate_record} and Figure \ref{Fig_9SAC_errorbox_record} confirm the effectiveness of the proposed approach in seismic UQ applications.


\subsection{Application to transmission tower structure}
\begin{figure}[H]
  \centering
  \includegraphics[scale=0.45] {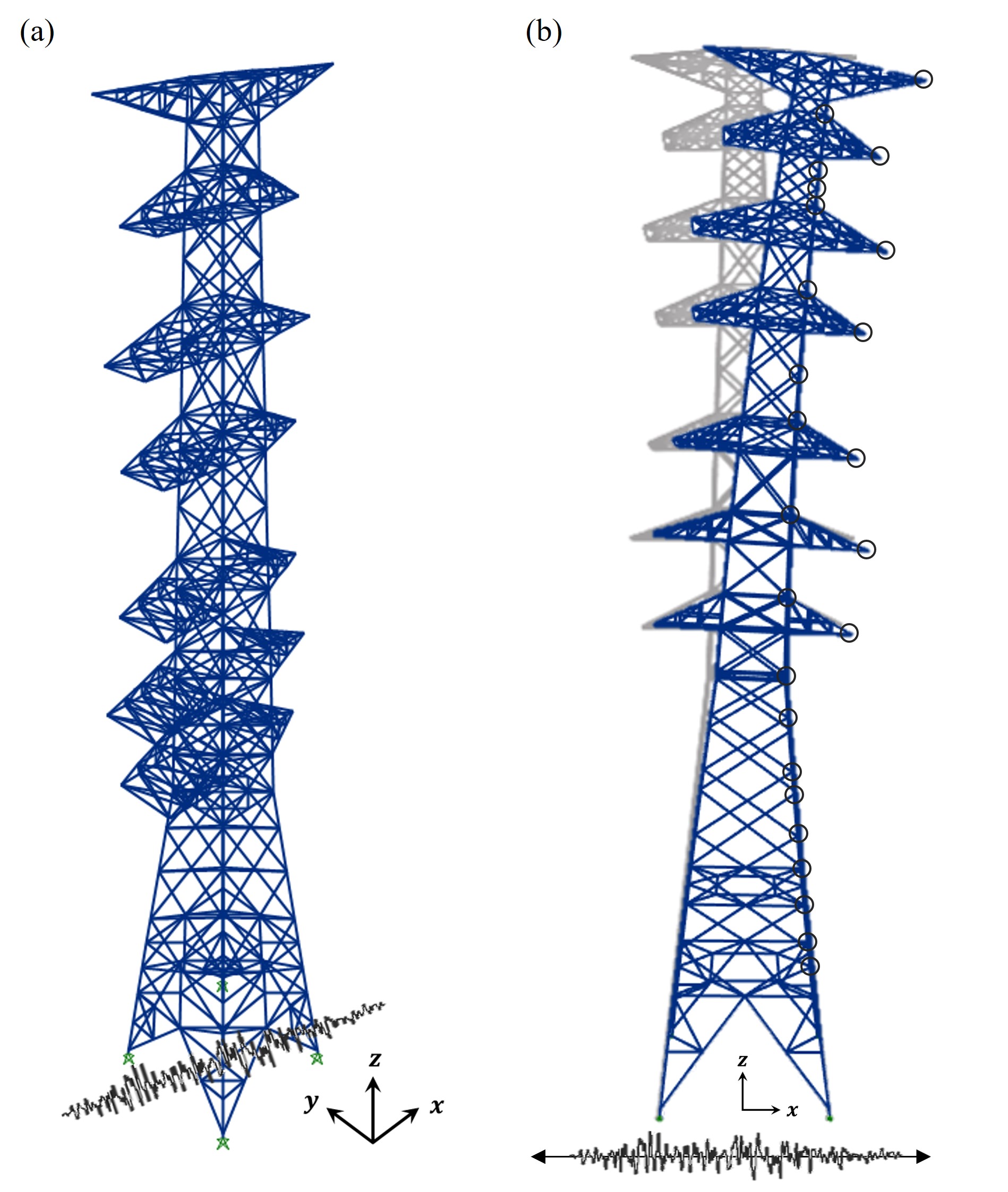}
  \caption{\textbf{A transmission tower structure: (a) perspective view and (b) front view}. {In figure (b), black circles denote the nodes where seismic responses, $\vect{Y}$, are obtained.}}
  \label{Fig_TT}
\end{figure}

\noindent This section evaluates the seismic responses of a high-rise transmission tower subjected to transverse seismic loads. Figure \ref{Fig_TT} illustrates the finite element model of the tower, constructed using SAP2000 for NLRHA \cite{park2016seismic}. The model includes continuous panels with cross-arms and a rigidly fixed foundation at the base. The height of the tower is 86.6 m. The structural model treats the modulus of elasticity $E$ and yield strength $\sigma_y$ of three different steel types as random variables in $\vect{X}_s$. Their distributions are listed in Table \ref{Tab_TTrvs}. The peak displacements at designated heights of the tower are of interest, represented as $\vect{Y} = \{Y_1, \dots, Y_{25}\} \in \mathbb{R}^{25}$.

\begin{figure}[H]
  \centering
  \includegraphics[scale=0.50] {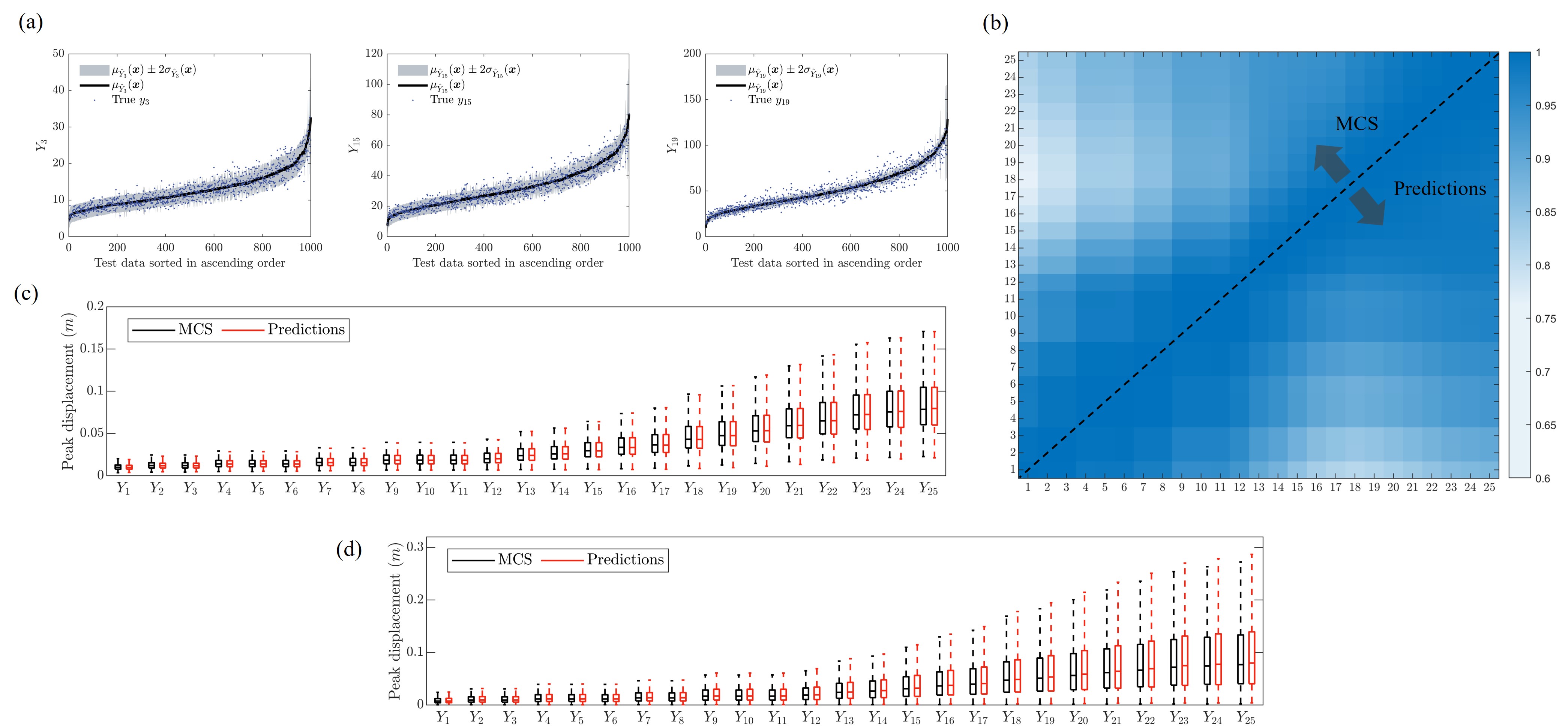}
  \caption{\textbf{Uncertainty quantification in seismic responses for the transmission tower example: (a) scatter plots, (b) correlation coefficient matrix, and (c) median and interquartile ranges of responses under artificially generated ground motions; (d) median and interquartile ranges of responses under real ground motion records}. {The proposed simulator is trained using $600$ samples. In figure (a), RMSE values are computed as $0.1396$, $0.1093$, and $0.0901$ for $Y_3$, $Y_{15}$, and $Y_{19}$, respectively. In figure (b), the axis labels denote the respective response variables from $\vect{Y}=\{Y_1,...,Y_{25}\}$.}}
  \label{Fig_TT_UQ0}
\end{figure}

Utilizing a training dataset of 600 samples, the optimal reduced dimension is determined to be $d=31$, with a threshold $\varepsilon^d_t=0.001$. The analysis includes both synthetically generated ground motions and real ground motion records, as discussed in Section \ref{3SAC}. Figure \ref{Fig_TT_UQ0} presents the UQ results from the proposed stochastic simulator, compared with results from direct MCS. The results again confirm the effectiveness of the proposed simulator in the high-dimensional seismic UQ problem involving computationally expensive NLRHAs.

\begin{table}[H]
  \caption{\textbf{Random variables associated with structural model for the transmission tower example}.}
  \label{Tab_TTrvs}
  \centering
  \begin{tabular}{c c c c}
    \toprule
    Random variables & Mean & Standard deviation & Distribution \\
    \midrule
    $E_{S240}$ (Mpa) & 200000 & 6000 & Lognormal \\
    $E_{S250}$ (Mpa) & 200000 & 6000 & Lognormal \\
    $E_{S335}$ (Mpa) & 200000 & 6000 & Lognormal \\
    $\sigma_{y,S240}$ (Mpa) & 240 & 12 & Lognormal \\
    $\sigma_{y,S250}$ (Mpa) & 250 & 12.5 & Lognormal \\
    $\sigma_{y,S335}$ (Mpa) & 335 & 16.75 & Lognormal \\
    \bottomrule
  \end{tabular} \\
\end{table}

\section{Conclusions} \label{Conclusion}

The proposed stochastic simulator-based uncertainty quantification method aims to quantify the variability of seismic responses, addressing the propagation of diverse sources of input uncertainties. The challenges are predominantly characterized by the complex, high-dimensional nature of ground motion uncertainties and the considerable computational demand necessitated by repeated NLRHAs. By leveraging physics-based dimensionality reduction, which exploits the intrinsic physical properties of ground motions, combined with a multivariate conditional distribution model, this method significantly enhances the scope of the existing dimensionality reduction-based surrogate modeling method (DR-SM) to more comprehensively tackle seismic UQ challenges. The performance of the proposed method is validated through its application to three different finite element building structures, demonstrating its capabilities to: (1) accurately predict multivariate seismic responses and (2) effectively quantify uncertainties, including the correlation structures among responses. Each scenario, involving both synthetic and real ground motion data, confirms the simulator's extensive applicability in seismic engineering practices.

While this paper demonstrates the effectiveness of the simulator for multi-output problems with up to $25$ dimensions, the extension to higher-dimensional outputs can be of particular interest. In engineering applications, where responses across multiple degrees of freedom can be highly correlated, methodologies such as proper orthogonal decomposition \cite{chatterjee2000introduction} and load-dependent Ritz vectors \cite{leger1987generation,wang2014multiple} can represent the high-dimensional output space using a few critical modes. Combining these methods with the proposed stochastic simulator has the potential to develop effective surrogate modeling approaches capable of addressing both high-dimensional input and output. Although the proposed stochastic simulator is demonstrated through global surrogate modeling for seismic UQ analysis, its accuracy in the tail regions of the distribution, which is crucial for seismic risk assessment, remains limited. This limitation arises from the use of LHS in designing the training points. Performance could be enhanced by incorporating stratified sampling \cite{neyman1992two,arunachalam2023efficient} and active learning \cite{echard2011ak,wang2020novel,kim2024efficient} techniques. Therefore, further research is warranted to explore the potential of the proposed method, particularly through integrating the stochastic simulator with stratified sampling and active learning techniques for rare event simulations.

\section*{Acknowledgement}
This research was supported by the Pacific Earthquake Engineering Research Center grant NCTRZW. The first author was also supported by the Basic Science Research Program through the National Research Foundation of Korea (NRF) funded by the Ministry of Education (RS-2024-00407901). The authors are grateful to Dr.~Taeyong Kim for providing the finite element models used in this paper. 





\bibliography{SeismicUQ_DRSM}


\end{document}